\documentclass[12pt]{article}
\pdfoutput=1
\usepackage[a4paper]{geometry}
\usepackage{jheppub, amsmath,amssymb,amsfonts,amsxtra, mathrsfs, makeidx,graphics,graphicx,amsthm,epsfig, ytableau,bm,longtable,float, color,tikz,mathtools,xfrac,footnote,rotating, lscape}
\usetikzlibrary{positioning}
\usetikzlibrary{chains}
\usetikzlibrary{arrows,fit,decorations.pathreplacing}
\tikzstyle{every picture}+=[remember picture]
\tikzstyle{na} = [baseline=-.5ex]

\usepackage{bbm}

\newcommand{\eg}{\textit{e.g.}}

\newcommand{\ie}{\textit{i.e.}}

\numberwithin{equation}{section}

\newcommand{\nn}{\nonumber}

\newcommand{\be}{\begin{equation}} \newcommand{\ee}{\end{equation}}
\newcommand{\bea}{\begin{equation} \begin{aligned}} \newcommand{\eea}{\end{aligned} \end{equation}}

\def\tilde{\widetilde}

\def\rt2{\sqrt{2}}

\def\det{\mathop{\rm det}}

\def\Tr{\mathop{\rm Tr}}

\def\CH{{\cal H}}

\def\CM{{\cal M}}
\def\CN{{\cal N}}

\def\CW{{\cal W}}

\def\1{{\ds 1}}

\newcommand{\cG}{\mathcal{G}}
\newcommand{\cH}{\mathcal{H}}

\newcommand{\cN}{\mathcal{N}}

\newcommand{\bZ}{\mathbb{Z}}

\def\repa{\raise4pt\hbox{$\square$}\mkern-14mu\raise-4pt\hbox{$\square$}}
\def\repab{\overline{\raise4pt\hbox{$\square$}\mkern-14mu\raise-4pt\hbox{$\square$}\mkern-1mu}}

\def\smileface{\ensuremath{\hbox{\large$\bigcirc$}\mkern-15mu\raise-1pt\hbox{\scriptsize$\smallsmile$}%
\mkern-10mu\raise4pt\hbox{..}\mkern4mu}}
\def\frownface{\ensuremath{\hbox{\large$\bigcirc$}\mkern-15mu\raise-1pt\hbox{\scriptsize$\smallfrown$}%
\mkern-10mu\raise4pt\hbox{..}\mkern4mu}}

\newcommand{\ba}{\begin{array}}
\newcommand{\ea}{\end{array}}
\newcommand{\bi}{\begin{itemize}}
\newcommand{\ei}{\end{itemize}}
\def\vec#1{\bm{#1}}
\def\bea#1\eea{\allowdisplaybreaks \begin{align}#1\end{align}}
 \newcommand{\ben}{\begin{enumerate}}
\newcommand{\een}{\end{enumerate}}
\newcommand{\bean}{\begin{eqnarray*}}
\newcommand{\eean}{\end{eqnarray*}}
\newcommand{\eref}[1]{(\ref{#1})}

\newcommand{\PE}{\mathop{\rm PE}}
\newcommand{\PL}{\mathop{\rm PL}}

\newcommand{\BC}{\mathbb{C}}

\newcommand{\BZ}{\mathbb{Z}}

\newcommand{\comment}[1]{}

\newcommand{\tI}{\widetilde{I}}
\newcommand{\Sym}{\mathrm{Sym}}

\def\skipper{\hskip.5em\relax}
\def\Esix#1#2#3#4#5#6{%
{\text{\small$ \left[ \begin{array}{c@{\skipper}c@{\skipper}c@{\skipper}c@{\skipper}c@{\skipper}c}
&&#2&& \\
#1 &  #3 &  #4 & #5 & #6
\end{array} \right]$}}}
\def\Eseven#1#2#3#4#5#6#7{%
{\text{\small$ \left[ \begin{array}{c@{\skipper}c@{\skipper}c@{\skipper}c@{\skipper}c@{\skipper}c}
&&#2&&& \\
#1 &  #3 &  #4 & #5 & #6&#7
\end{array} \right]$}}}
\def\Eeight#1#2#3#4#5#6#7#8{%
{\text{\small$\begin{array}{c@{\skipper}c@{\skipper}c@{\skipper}c@{\skipper}c@{\skipper}c@{\skipper}c@{\skipper}c}
&&#2&&&& \\
#1 &  #3 & #4  & #5 & #6 & #7 &  #8
\end{array}$}}}

\newfloat{sourcecode}{thp}{lop}
\floatname{sourcecode}{SC}

\title{Instanton Operators and the Higgs Branch at Infinite Coupling}
\author[a]{Stefano Cremonesi,}
\author[b]{Giulia Ferlito,}
\author[b]{Amihay Hanany,}
\author[c]{Noppadol Mekareeya}

\affiliation[a]{Department of Mathematics, King's College London, \\
The Strand, London WC2R 2LS, United Kingdom}
\affiliation[b]{Theoretical Physics Group, Imperial College London, \\
Prince Consort Road, London, SW7 2AZ, UK}
\affiliation[c]{Theory Division, Physics Department, CERN, \\CH-1211, Geneva 23, Switzerland}

\emailAdd{stefano.cremonesi@kcl.ac.uk}
\emailAdd{giulia.ferlito11@imperial.ac.uk}
\emailAdd{a.hanany@imperial.ac.uk}
\emailAdd{noppadol.mekareeya@cern.ch}

\preprint{
{\small
\begin{flushright}
IMPERIAL-TP-15-AH-03\\
CERN-PH-TH-2015-115
\end{flushright}
}
}

\abstract{The richness of 5d $\mathcal{N}=1$ theories with a UV fixed point at infinite coupling is due to the existence of local disorder operators known as instanton operators. By considering the Higgs branch of $SU(2)$ gauge theories with $N_f \leq 7$ flavours at finite and infinite coupling, we write down the explicit chiral ring relations between instanton operators, the glueball superfield and mesons. Exciting phenomena appear at infinite coupling: the glueball superfield is no longer nilpotent and the classical chiral ring relations are quantum corrected by instanton operators bilinears. We also find expressions for the dressing of instanton operators of arbitrary charge. The same analysis is performed for $USp(2k)$ with an antisymmetric hypermultiplet and pure $SU(N)$ gauge theories.}

\begin{document}
\maketitle

\section{Introduction}
The dynamics of five-dimensional supersymmetric gauge theories has many interesting features.  From the Lagrangian perspective these field theories are not renormalisable.  However, by using string theoretic methods along with field theory analysis, it was demonstrated that a number of such field theories can be considered as flowing from certain non-trivial superconformal field theories in the ultraviolet (UV) \cite{Seiberg:1996bd, Douglas:1996xp, Morrison:1996xf, Intriligator:1997pq}.  
Such UV fixed points at infinite gauge coupling may exhibit an enhancement of the global symmetry.  In particular, in the seminal work \cite{Seiberg:1996bd}, it was pointed out that the UV fixed point of 5d $\CN=1$ $SU(2)$ gauge theory with $N_f \leq 7$ flavours exhibits $E_{N_f+1}$ flavour symmetry, which enhances from the global symmetry $SO(2N_f) \times U(1)$ apparent in the Lagrangian at finite coupling.   Since then a large class of five dimensional supersymmetric field theories have been constructed using webs of five-branes \cite{Aharony:1997ju, Aharony:1997bh, DeWolfe:1999hj} and the enhancement of the global symmetry of these theories has been studied using various approaches, including superconformal indices \cite{Kim:2012gu, Rodriguez-Gomez:2013dpa, Bergman:2013koa, Bergman:2013ala,Taki:2013vka, Bergman:2013aca, Hwang:2014uwa, Zafrir:2014ywa, Bergman:2014kza, Zafrir:2015uaa, Hayashi:2015fsa, Yonekura:2015ksa}, Nekrasov partition functions and (refined) topological string partition functions \cite{Bao:2011rc, Bashkirov:2012re, Iqbal:2012xm, Bao:2013pwa, Hayashi:2013qwa, Hayashi:2014wfa, Mitev:2014jza, Kim:2014nqa, Hayashi:2015xla}.

In five dimensions, instantons are particles charged under the the $U(1)$ global symmetry associated  with the topological conserved current $J = \frac{1}{8 \pi^2} \Tr * (F \wedge F)$; this global symmetry is denoted by $U(1)_I$ in the rest of the paper.  In the UV superconformal field theory, the instanton particles are created by local operators known as {\it instanton operators}, that insert a topological defect at a spacetime point and impose certain singular boundary conditions on the fields %(see \eg~ 
\cite{Lambert:2014jna, Rodriguez-Gomez:2015xwa, Tachikawa:2015mha}.
  These operators play an important role in enhancing the global symmetry of the theory.  For 5d $\CN=1$ field theories at infinite coupling, it was argued that instanton operators with charge $I=\pm 1$, form a multiplet under the supersymmetry and flavour symmetry \cite{Tachikawa:2015mha}.  In 5d $\CN=2$ Yang-Mills theory with simply laced gauge group, it is believed that the instanton operators constitute the Kaluza-Klein tower that enhances the Poincar\'e symmetry and provides the UV completion by uplifting this five dimensional theory to the $6d$ $\CN=(2,0)$ CFT \cite{Douglas:2010iu, Lambert:2010iw, Lambert:2014jna}.

Standard lore says that the Higgs branch of theories with 8 supercharges in dimensions 3 to 6 are classically exact, and do not receive quantum corrections. In 5 dimensions, this statement turns out to be imprecise, and should be corrected. In fact, one of the main points of the paper, is that there are three different regimes, given by 0, finite, and infinite gauge coupling. The hypermultiplet moduli space, which we always refer to as the Higgs branch, turns out to be different in each of these regimes, and hence our analysis corrects and sharpens the standard lore. The main goal of the paper is to understand how, at infinite coupling, instanton operators correct the chiral ring relations satisfied by the classical fields at finite coupling.

In order to perform such an analysis we start from the known Higgs branch at infinite coupling and write the Hilbert series of such a moduli space for various 5d $\CN=1$ theories. We mostly focus on the $SU(2)$ gauge theories with $N_f$ flavours, for which string theory arguments show that the Higgs branch at infinite coupling is the reduced moduli space of one $E_{N_f+1}$ instanton on $\BC^2$  \cite{Seiberg:1996bd, Aharony:1997bh}. The Hilbert series counts the holomorphic functions that parametrise the Higgs branch, graded with respect to the Cartan subalgebra of the (enhanced) flavour symmetry and the highest weight of the $SU(2)$ $R$-symmetry of the theory: 
\bea
H(t,y) = \mathrm{Tr}_\cH \Big(t^{2R} y_A^{H_A} \Big)~,
\eea  
where $\cH$ is the Hilbert space of chiral operators of the SCFT, $R$ the $SU(2)_R$ isospin and $H_A$ the Cartan generators of the enhanced global symmetry. 

Such a Hilbert series can then be expressed in terms of the global symmetry of the theory at finite coupling --- the latter is a subgroup of the enhanced symmetry at infinite coupling:  
\bea
H(t, y(x,q)) = \mathrm{Tr}_\cH \Big(t^{2R} q^I x_a^{H_a} \Big)~,
\eea
where $I$ is the topological charge and $H_a$ the Cartan generators of the $SO(2N_f)$ flavour symmetry.
This decomposition allows us to extract the contributions of the classical fields and the instanton operators to the Higgs branch chiral ring and explicitly write down the relations they satisfy.

The paper is organised as follows. In section \ref{sec:SU2} we study the Higgs branch of $SU(2)$ gauge theories with $N_f\leq 7$ flavours, spell out the relations in the chiral ring in terms of mesons, glueball and instanton operators, and discuss the dressing of instanton operators. We generalise the analysis to pure $USp(2k)$ Yang-Mills theories with an antisymmetric hypermultiplet in sections \ref{sec:USp4} and \ref{sec:USp2k}, and to pure $SU(N)$ Yang-Mills in section \ref{sec:SUN}. We close the paper with a discussion of our results and an outlook in section \ref{sec:Discussion}. Several technical results are relegated to appendices.

\section{$SU(2)$ with $N_f$ flavours: one $E_{N_f+1}$ instanton on $\BC^2$}\label{sec:SU2}

The dynamics of 5d $\CN=1$ $SU(2)$ gauge theory with $N_f \leq 7$ flavours was studied in detail in \cite{Seiberg:1996bd}. In there it was argued that, despite being power counting non-renormalisable, these theories possess strongly interacting UV fixed points. Moreover a classification was proposed where the global symmetry, which at finite coupling is $SO(2N_f) \times U(1)_I$, with $U(1)_I$ the global symmetry associated with a topologically conserved current,  enhances to $E_{N_f+1}$, where $\widetilde{E}_1=U(1),E_1=SU(2)$, $E_2 = SU(2) \times U(1)$, $E_3= SU(3) \times SU(2)$, $E_4= SU(5)$, $E_5=SO(10)$ and $E_6, E_7, E_8$ are the usual exceptional symmetries.

The analysis presented in this paper focuses on how the Higgs branch of these 5d theories changes along the RG flow. In particular we take care in distinguishing three different regimes for these theories, the operators that contribute to the chiral ring on the Higgs branch% 
\footnote{Even though we discuss theories with minimal $\mathcal{N}=1$ supersymmetry (that is $8$ Poincar\'e supercharges) in 5 dimensions, we are interested in the chiral ring as defined in terms of a subsuperalgebra with $4$ supercharges, and the Higgs branch as a complex algebraic variety. We therefore use 4d $\cN=1$ notation and terminology throughout this paper. Even though this formalism is not consistent with Poincar\'e supersymmetry in five dimensions, it is necessary to discuss chiral operators and holomorphic functions on the Higgs branch.}
and the defining equations that these operators satisfy:
\begin{itemize}
 \item In the classical regime, where fermions are neglected, these 5d theories have the usual Higgs branch which is just given by $\widetilde{\CM}_{1, SO(2N_f)}$, the centred (or \emph{reduced}) moduli space of  one $SO(2N_f)$ instanton. The gauge invariant operators that generate this space are mesons $M^{ab}$, constructed out of chiral matter superfields in the bifundamental of the $SU(2)$ gauge group and $SO(2N_f)$ flavour group.
 The relations that these generators satisfy on the moduli space can be extrapolated from its description as the minimal nilpotent orbit of $SO(2N_f)$ \cite{kronheimer1990}. They are the usual Joseph relations \cite{Joseph1976} and their transformation properties can be read off from the decomposition of the second symmetric product of the adjoint, the representation in which the generator transforms. Let $V(\theta)$ denote the adjoint representation. The decomposition
\bea \label{generalJoseph}
\Sym^2 V(\theta) = V(2 \theta) + \mathfrak{I}_2
\eea
prescribes that the relations transform in the representation $\mathfrak{I}_2$.

For $SO(2N_f)$
\bea \label{JosephForSO2N}
\mathfrak{I}_2=\Sym^2 [1,0,...]+\wedge^4[1,0,...]~.
\eea
We can construct these representations from the adjoint mesons $M^{ab}$ as follows. Take $M$ to be an antisymmetric $2N_f \times 2N_f$ matrix,  $M^{ab}=-M^{ba}$, $a,b=1,..,2N_f$. Then the two terms of \eref{JosephForSO2N} correspond respectively to:
\bea
M^2&=0 \label{ExplicitJosephForSO2N-Symm}\\
M^{[ab}M^{cd]}&=0~ \label{ExplicitJosephForSO2N-Antisymm}.
\eea
We call the last equation the rank 1 condition, since for an antisymmetric matrix it is equivalent to the vanishing of all degree 2 minors.

 \item When the coupling is finite, one needs to take into account the contribution from the gaugino sector. In particular, the glueball superfield $S$, which is a chiral superfield bilinear in the gaugino superfield $\mathcal{W}$, is now no longer suppressed and will \emph{de jure} appear in the chiral ring. This operator satisfies a classical relation in the chiral ring as in four dimensions \cite{Cachazo:2002ry}, namely
 \bea \label{CRrelnSU2}
 S ^2=0~,
 \eea 
hence $S$ is the only extra operator that one needs to consider at finite coupling. At first sight it might seem counterintuitive that $S$ contributes to the Higgs branch as it is a bilinear in the vector multiplet. In fact in 5d the Higgs branch is the \emph{only} complex branch of the full moduli space. As such, any chiral operator, and in particular the glueball superfield $S$, belongs to the class of Higgs operators. This will become even clearer later, when we recover the finite coupling Higgs branch from the one at infinite coupling.

Geometrically we interpret the operator $S$ as generating a 2-point space, which by a slight abuse of notation we denote by $\mathbb{Z}_2$. Algebraically the Hilbert series for this space is simply written as 
\bea \label{HSforSfiniteCoupl}
HS(\mathbb{Z}_2; t)=1+t^2
\eea
where 1 signifies the identity operator and the $t^2$ term is associated to the quadratic operator $S$. The fugacity $t$ grades operators by their $SU(2)_R$ representation and the normalisation is chosen so that the power %signifies 
is twice the isospin. The meson $M^{ab}$ and the glueball superfield $S$ obey the %following relation 
chiral ring relation \cite{Cachazo:2002ry, Argurio:2003ym}
\bea \label{SMreln}
S M^{ab}=0~.
\eea
This signifies that the spaces $\widetilde{\CM}_{1, SO(2N_f)}$ and $\mathbb{Z}_2$ intersect only at the origin.

From an algebraic perspective, when two moduli spaces $X$ and $Y$ intersect, the Hilbert series of their union is given by the surgery formula
\bea \label{surgeryformula}
H_{X \cup Y}=H_X+H_Y-H_{X \cap Y}~,
\eea
where the subtraction is done to avoid double counting \cite{Hanany:2006uc}. Thus, when $\mathbb{Z}_2$ is glued to $\widetilde{\CM}_{1, SO(2N_f)}$, the net effect on the Hilbert series is simply that of adding a $t^2$ to the Hilbert series of $\widetilde{\CM}_{1, SO(2N_f)}$.

The plethystic logarithm%
\footnote{The plethystic logarithm of a multivariate function $f(x_1,...,x_n)$ such that $f(0,...,0)=1$ is
\bea
PL[f(x_1,...,x_n)]=\sum^\infty_{k=1} \frac{1}{k} \mu(k) \log f(x^k_1,...,x^k_n)
\eea
where $\mu(k)$ is the Moebius function.  The plethystic logarithm of the Hilbert series encodes generators and relations of the chiral ring.}
of this newly obtained expression is interesting: it shows that at order $t^4$ there are two extra relations compared to the classical regime, one transforming in the singlet and one transforming in the adjoint of $SO(2N_f)$. The singlet relation is \eref{CRrelnSU2}. For the adjoint relation the only possible extra operator that one can construct in such a representation is $S M^{ab}$. The adjoint relation is then precisely \eref{SMreln}.

\item At infinite coupling, the moduli space is a different space altogether. Instanton operators, %and anti-instanton operators, $I$ and $\widetilde{I}$ respectively, 
carrying charge under $U(1)_I$, contribute to the chiral ring and are responsible for prompting symmetry enhancement: the Higgs branch in this regime becomes isomorphic to the reduced moduli space $\widetilde{\CM}_{1, E_{N_f+1}}$ of one $E_{N_f+1}$ instanton on $\BC^2$ \cite{Seiberg:1996bd}. In order for this to happen a crucial event on the chiral ring takes place: instanton and anti-instanton operators $I$ and $\widetilde{I}$ of $U(1)_I$ charge $\pm 1$ correct the relation \eref{CRrelnSU2}.%
\footnote{We call the instanton operator $\widetilde{I}$ of topological charge $-1$ ``anti-instanton operator'', even though it is mutually BPS with the positively charged instanton operator $I$.}

This is the most dramatic dynamical mechanism happening at infinite coupling: the operator $S$ is no longer a nilpotent bilinear in the vector multiplet and it becomes, for all intents and purposes, a chiral bosonic operator on the Higgs branch. The contribution of $S$ to the chiral ring will no longer amount to \eref{HSforSfiniteCoupl}, but instead an infinite tower of operators will appear generating a factor $(1-t^2)^{-1}$ in the Hilbert series.
\end{itemize}
 
The purpose of this paper is to explore these statements quantitatively for known cases of UV-IR pairs of theories. We do this as follows. We start from the UV theory at infinite gauge coupling, which has $E_{N_f+1}$ symmetry acting on the hypermultiplet moduli space. As soon as the dimensionful gauge coupling becomes finite, a term is added to the scalar potential which is proportional to the norm squared of the moment maps of the broken symmetries in the breaking $E_{N_f+1} \rightarrow SO(2N_f) \times U(1)_I$. Consequently, the broken moment maps must vanish on the Higgs branch of the theory at finite  coupling. In terms of the chiral ring, this sets to zero the instanton operators $I$ and $\widetilde{I}$.% 
\footnote{Although this argument applies to most of the theories we study in this paper, it is in general not useful for theories where instanton operators have $SU(2)_R$ spin higher than 1, \emph{e.g.} as in section \ref{sec:SUN}.}

Computationally, one starts with the Hilbert series of the reduced one $E_{N_f+1}$ instanton moduli space written in terms of representations of $E_{N_f+1}$ \cite{Benvenuti:2010pq} and decomposes them into representations of $SO(2N_f) \times U(1)_I$. For all theories of our interest, the Hilbert series after this decomposition admits a very simple expression in terms of the highest weight generating function \cite{Hanany:2014dia}.  This allows us to analyse the generators of the moduli space in terms of instanton operators and classical fields, and in many cases the relations between such generators are sufficiently simple to be written down explicitly.

\subsection{$E_0$}
The $E_0$ theory is the trivial case. There is no hypermultiplet moduli space. Consequently the Hilbert series for this theory is just given by 1, corresponding to the identity operator. The theory has no RG flow. Its interest lies in it being the limiting case of all the theories we consider in this section since none of the operators ($M$,$S$,$I, \widetilde{I}$) makes an appearance.

\subsection{$N_f =0$}
A pure $SU(2)$ SYM theory with $\CN=1$ supersymmetry in 5d can be obtained by flowing from two UV fixed points which have different global symmetry. The existence of these two theories is dictated by a discrete $\theta$ parameter taking value in $\pi_4(Sp(1))=\mathbb{Z}_2$ \cite{Douglas:1996xp}. For the non-trivial element the global symmetry at infinite coupling is $\widetilde{E}_1=U(1)$ whilst for the identity element the global symmetry is $E_1=SU(2)$. 

\subsubsection{The $\widetilde{E}_1$ theory }

For the theory with $\theta=\pi$ no enhancement of the global symmetry occurs: the global symmetry at finite and infinite coupling is the instanton charge symmetry $U(1)_I$.
Here instanton operators are absent and the generator of the moduli space is just $S$ obeying $S^2=0$, both at infinite and finite coupling. The moduli space generated by this operator is simply $\mathbb{Z}_2$. Classically the moduli space is trivial.

\subsubsection{The $E_1$ theory }
For the theory associated to the trivial element of the $\mathbb{Z}_2$ valued $\theta$ parameter the $U(1)_I$ topological symmetry is enhanced to $SU(2)$ by instanton operators at infinite coupling. In this regime the Higgs branch of the theory is isomorphic to the reduced moduli space of one-$SU(2)$ instanton $\widetilde{\CM}_{1, SU(2)}$, which is the orbifold $\mathbb{C}^2 / \mathbb{Z}_2$. This theory is the prototypical example of the class we study. Since there is no flavour symmetry, we can understand the three regimes by means of simple physical arguments.

As we flow away from the UV fixed point, the Higgs branch is lifted and its only remnant is a discrete $\mathbb{Z}_2$ space generated by $S$. Classically, even this contribution can be neglected and the Higgs branch is completely absent. This is a remarkable effect whereby from no Higgs branch in the classical regime a full Higgs branch opens up at infinite coupling.

Algebraically we start from the Hilbert series for $\mathbb{C}^2 / \mathbb{Z}_2$ and decompose it in representations of $U(1)_I$ so that we can identify the contribution from instanton operators, as well as the finite coupling chiral operators, and their relations. 

The Hilbert series for $\mathbb{C}^2 / \mathbb{Z}_2$ can be written as
\bea
H[\widetilde{\CM}_{1, SU(2)}](t; x) = \sum_{n=0}^\infty [2n]_x t^{2n} = \frac{1-t^4}{(1-t^2 x^2)(1-t^2)(1-t^2 x^{-2})}~,
\eea
where $t$ is the fugacity for the $SU(2)_R$ symmetry, $x$ is the fugacity for the $SU(2)$ global symmetry acting on $\mathbb{C}^2 / \mathbb{Z}_2$, and $[2n]_x$ stands for the character, as a function of $x$, of the representation of $SU(2)$ with such a Dynkin label.
Identifying the Cartan subalgebra of the $SU(2)$ symmetry with $U(1)_I$, we obtain
\be \label{HS1-SU2}
\begin{split}
H[\widetilde{\CM}_{1, SU(2)}](t; q^{1/2}) &=  \frac{1-t^4}{(1-t^2 q)(1-t^2)(1-t^2 q^{-1})} \\
&= \frac{1}{1-t^2} \sum_{j=-\infty}^\infty t^{2|j|} q^j~.
\end{split}
\ee

\subsubsection{The generators and their relations}
Eq. \eref{HS1-SU2} has a natural interpretation in terms of operators at infinite coupling:
\bi
\item Each term in the sum $t^{2|j|} q^j$ corresponds to an instanton operator $I_{+|j|}$ for $j>0$ and an anti-instanton operator $I_{-|j|}$ for $j<0$ that is the highest weight state of the $SU(2)_R$ representation with highest weight $2|j|$.% 
\footnote{Notice how the $SU(2)_R$ spin of an instanton operator of charge $\pm j$ is $|j|$. Whilst we can easily extract the $SU(2)_R$ spin as a function of instanton number, it is not clear how to do so for the representation under the global symmetry, as will be seen for the cases with higher number of flavours.} %\\
 $q$ is the fugacity for the instanton number $U(1)_I$. The plethystic logarithm of the Hilbert series shows that the instanton operator $I_{+|j|}$ is generated by the charge 1 operator  $I_{+1} \equiv I$ through the relation $I_{+|j|}=(I)^j$. Similarly $I_{-|j|}=(\widetilde{I})^j$ where $\widetilde{I} \equiv I_{-1}$.
\item The tower of operators generated by $S$ can be identified with the factor $(1-t^2)^{-1}$. This enhancement in the number of operators constructed from powers of $S$ is crucial: at infinite coupling $S$ is a full-on operator on the Higgs branch and, together with the instanton and anti-instanton operators $I$, $\widetilde{I}$, forms a triplet of the $SU(2)$ that generates $\mathbb{C}^2 / \mathbb{Z}_2$.  
\ei
From this form of the Hilbert series we can also give another interpretation to the Higgs branch at infinite coupling. Instanton operators on the Higgs branch in 5d $\mathcal{N}=1$ theories play a similar role to monopole operators in 3d $\cN=4$ \cite{Cremonesi:2013lqa} and $\cN=2$ theories \cite{Hanany:2015via,Cremonesi:2015dja}: in this sense \eref{HS1-SU2} can be interpreted as the space of \emph{dressed} instanton operators, where the factor $\frac{1}{1-t^2}$ is the dressing from the operator $S$ and it is freely generated.

The numerator in the rational function of \eref{HS1-SU2} signifies a relation quadratic in the operators which can only be given by
\bea
S^2=I \widetilde{I} ~,
\eea
the defining equation for $\mathbb{C}^2 / \mathbb{Z}_2$.

At finite coupling, where $I, \widetilde{I}=0$, we recover the known chiral ring relation \eref{CRrelnSU2}, i.e. the nilpotency of the operator $S$. As we have explained, the only remnant of $\mathbb{C}^2 / \mathbb{Z}_2$ is a residual $\mathbb{Z}_2$ generated precisely by $S$.

Classically, we can set $S=0$ and lift the Higgs branch entirely. 

\subsection{$N_f =1$}
For $N_f=1$ and $N_f=2$ the infinite coupling Higgs branch is the moduli space of one instanton for a product gauge group. In such cases the moduli space is given by the union of the one instanton moduli space for each factor. 
For the case of $N_f=1$, i.e $E_2 = SU(2) \times U(1)$, the Higgs branch at infinite coupling is thus the union of the one $SU(2)$ and the one $U(1)$ instanton moduli spaces.

For the $U(1)$ instanton moduli space, there are two possible ADHM constructions that one may consider: (1) $USp(2)$ gauge theory with one flavour, and (2) $U(1)$ gauge theory with one flavour.  As analysed below, the Higgs branch of the former is $\BZ_2$ whereas the Higgs branch of the latter is a point.   A priori it might not be apparent which option is the correct one but consistency with the finite coupling regime points out that the right choice is the former.  We provide an independent argument below.
%The global structure of $E_2$ turns out to be important. Indeed, treating the abelian factor as a $U(1)$ or as an $SO(2)$ group yields two moduli spaces which differ by one point. A priori one would not be able to select one over the other but consistence with the classical regime moduli space will point out the correct choice.

Let us begin with the first option.  The Higgs branch of the ADHM gauge theory given by $USp(2)$ with one flavour describes the moduli space of one $SO(2)$ instanton.\footnote{To be precise, the flavour symmetry of the quiver gauge theory is $O(2)$, not $SO(2)$. (We thank the referee for raising this point.) However the moduli space of instantons in question is insensitive to the difference between the two groups.} There is only one operator in the chiral ring, $P$, subject to a quadratic nilpotency relation, $P^2=0$. The moduli space of one $SO(2)$ instanton is thus $\BZ_2$.\footnote{Note that as rings $\mathbb{C}[P] / \langle P^2 \rangle \neq \mathbb{C}[P] / \langle P \rangle$.}

On the other hand, one may consider a $U(1)$ gauge theory with one flavour, whose Higgs branch is often referred to as ``the moduli space of one $U(1)$ instanton''. The gauge invariant quantity is $Q \tilde{Q}$ but is set to zero by the F-terms. The moduli space is thus trivial: it consists of one point only rather than two.

The reduced moduli space $\widetilde{\CM}_{1, E_2}$ of one $E_2$ instanton is thus either isomorphic to the space $\mathbb{C}^2 / \mathbb{Z}_2 \cup \mathbb{Z}_2$ or to $\mathbb{C}^2 / \mathbb{Z}_2 \cup \{1\}$, depending on which of the above options is correct.

With the first option, the Hilbert series of $\widetilde{\CM}_{1,E_2}$ can be written using \eref{surgeryformula} as:
\be \label{HS1E_2}
\begin{split}
H[\widetilde{\CM}_{1,E_2}](t; x) & = H[\widetilde{\CM}_{1,SU(2)}]+H[\mathbb{Z}_2]-1 \\
& = \frac{1-t^4}{(1-x^2 t^2) (1-t^2) (1-x^{-2} t^2)} + t^2
\end{split}
\ee
where $H[\mathbb{Z}_2]=1+t^2$ is generated by $P$.  

%The last $t^2$ term and the three generators at $t^2$ from the first term in \eref{HS1E_2} saturate the four generators of $SU(2) \times U(1)$.

With the second option, the Hilbert series of $\widetilde{\CM}_{1,E_2}$ is
\be \label{HS1E_2b}
\begin{split}
H[\widetilde{\CM}_{1,E_2}](t; x) & = H[\widetilde{\CM}_{1,SU(2)}] \\
& = \frac{1-t^4}{(1-x^2 t^2) (1-t^2) (1-x^{-2} t^2)}~.
\end{split}
\ee

The generator of the $\mathbb{C}^2 / \mathbb{Z}_2$ factor is $\Phi^{ij}$, $i=1,2$, with $\Phi^{ij}=\Phi^{ji}$ and it obeys the quadratic nilpotency:
\be \label{c2modZ2only}
\Phi^{ij} \epsilon_{jk} \Phi^{kl}=0
\ee
where $\epsilon_{ij}$ is defined by its antisymmetry property and $\epsilon_{12}=1$.  

The extra generator, $P$, is there only in the case of a union of $\mathbb{C}^2 / \mathbb{Z}_2$ with a two point moduli space. In its presence, beside  \eref{c2modZ2only},  two further relations hold: 
\be \label{extraZ2_relns}
\begin{split}
P^2=0 \\
P \Phi^{ij}=0
\end{split}
\ee
\eref{c2modZ2only} is the usual Joseph relation for the $SU(2)$ minimal nilpotent orbit $\mathbb{C}^2 / \mathbb{Z}_2$.   The last equation encodes the fact that the two spaces, $\mathbb{C}^2 / \mathbb{Z}_2$ and $\mathbb{Z}_2$, only intersect at one point, the origin of the moduli space.

Let us proceed without making any assumption on whether $\widetilde{\CM}_{1,E_2}$ is given by $\mathbb{C}^2 / \mathbb{Z}_2\, \cup\, \mathbb{Z}_2$ or $\mathbb{C}^2 / \mathbb{Z}_2\, \cup \, \{1\}$. In the next subsection, we show that consistency with the finite coupling result tells us that the correct choice is the former.

\subsubsection{The generators and their relations}
The theory at finite coupling has a Higgs branch which is isomorphic to the union of $\widetilde{\CM}_{1,SO(2)}$ with $\mathbb{Z}_2$, the former generated by a meson, $M$, subject to a quadratic nilpotency and the latter by the glueball superfield $S$, itself quadratically nilpotent. The finite coupling chiral ring is thus defined by:
\bea \label{finite_1flavour}
&M^2 = S^2 =S M =0
\eea
where the last equation signifies that the two spaces, $\widetilde{\CM}_{1,SO(2)}$ generated by $M$ and $\mathbb{Z}_2$ generated by $S$, are orthogonal to each other and intersect only at the origin. Moreover since $\widetilde{\CM}_{1,SO(2)} \cong \mathbb{Z}_2$, the Higgs branch at finite coupling is given by $\mathbb{Z}_2 \cup \mathbb{Z}_2$.

The goal is to reproduce the set of equations \eref{finite_1flavour} from the ones at infinite coupling by setting the instanton operators appearing there to zero.
This can be achieved as follows. Decompose the generators $\Phi^{ij}$ of $\widetilde{\CM}_{1,E_2}$ by letting
\bea
&\Phi^{1 1} = I \\
&\Phi^{1 2} =M  \\
&\Phi^{2 2}=-\widetilde{I}
\eea
where $M$ is the $SO(2)$ mesonic operator and $I, \widetilde{I}$ are the instanton and anti-instanton operators respectively. The relation in \eref{c2modZ2only} can then be rewritten as:
\bea
M^2=I \widetilde I~.
\eea
It is clear that, by setting the instanton operators to zero, only one of the three equations in \eref{finite_1flavour} can be recovered for the finite coupling limit. However, if the  extra operator $P$ and the extra relations in \eref{extraZ2_relns} are also taken into account, the classical regime can be precisely recovered. To this avail, let $P$ be decomposed as: 
\bea
P= S- M~,  
\eea
{\it i.e.} a linear combination of the meson $M$ and the glueball $S$. Then \eref{c2modZ2only} and \eref{extraZ2_relns} together can be rewritten as:
\bea
&M^2 = \widetilde{I} I \\
&S ^2=\widetilde{I} I  \\
&S M=\widetilde{I} I  \\
&MI=SI  \\
&\widetilde{I}M =\widetilde{I}S~. 
\eea
This time, setting $I, \widetilde{I} = 0$, the finite coupling relations \eref{finite_1flavour} are finally recovered. 

In the classical regime, where we neglect the contribution from $S$, we recover the space $\mathbb{Z}_2$, the reduced moduli space of one $SO(2)$ instanton generated by $M$, such that $M^2=0$. 

This is the required consistency that we mentioned above: $\widetilde{\CM}_{1,E_2}$ is indeed $\mathbb{C}^2 / \mathbb{Z}_2\, \cup\, \mathbb{Z}_2$, the latter being given by the ADHM construction of $USp(2)$ with 1 flavour.

Let us provide a complementary argument based on symmetries that supports the identification of $\widetilde{\CM}_{1,E_2}$ with $\mathbb{C}^2 / \mathbb{Z}_2\, \cup \, \BZ_2$.  The ADHM construction for $U(1)$ with $N_f$ flavours provides the moduli space of $U(N_f)/U(1)$ instantons, which for $N_f=1$ corresponds to an empty symmetry group and thus a trivial moduli space. Furthermore, in the presence of a flavour symmetry, an $SU(2)_R$ spin-1 operator is a necessary requirement for the existence of a linear hypermultiplet containing the conserved current.   For a $U(1)$ gauge theory with $1$ flavour, there is no flavour symmetry and hence no associated generator.  Identifying $\widetilde{\CM}_{1,E_2}$ with $\mathbb{C}^2 / \mathbb{Z}_2\, \cup \{1 \}$,  there would be only three generators transforming in the adjoint representation of $SU(2)$ associated with $\BC^2/\BZ_2$ but no extra generator associated with the aforementioned $U(1)$ symmetry, as in \eref{HS1E_2b}.  On the other hand, for a $USp(2)$ gauge theory with 1 flavour, there is an $SO(2) \cong U(1)$ flavour symmetry; hence there is a generator at order $t^2$ associated with this symmetry.   We see that only when we identify $\widetilde{\CM}_{1,E_2}$ with $\mathbb{C}^2 / \mathbb{Z}_2\, \cup \BZ_2$  there are four generators transforming in the adjoint representation of the global symmetry $SU(2) \times U(1) \cong E_2$ as one can see explicitly in \eref{HS1E_2}.

\subsubsection{Expansion in the instanton fugacity }
It is instructive to rewrite \eref{HS1E_2} as an expansion in $q$, the $U(1)_I$ fugacity. Replacing $x$, the fugacity for $SU(2)$, by $q^{1/2}$ we have that:
\bea \label{qexpNf1}
H[\widetilde{\CM}_{1, E_2}](t; y, q^{1/2}) &=\frac{1}{(1-t^2)} \sum^{\infty}_{n=-\infty } q^{n}  t^{|2n| } + t^2~.
\eea
Hence a bare instanton operator with $U(1)_I$ charge $n$ is the highest weight state of the spin $|n|$ representation of the $SU(2)_R$ symmetry. For $n \neq 0$, the tower of states originating from the glueball $(1-t^2)^{-1}$, i.e the space $\mathbb{C}$, acts as a dressing for the instanton operators.  For $n = 0$, the dressing is a different space, due to the presence of an extra piece of the moduli space unaffected by instantons. It is in fact the space generated by $S$ and $M$, subject to the relations $SM=0$ and $M^2=0$, i.e $\mathbb{C} \cup \mathbb{Z}_2$.

\subsection{$N_f =2$}

The reduced moduli space of one $E_3=SU(3) \times SU(2)_A$ instanton%
\footnote{The subscript $A$ is used to differentiate from $SU(2)_B$ which is defined in the next paragraph.} is isomorphic to the union of two hyperK\"ahler cones, the reduced moduli space of one $SU(3)$ instanton, $\widetilde{\CM}_{1, SU(3)}$, and the reduced moduli space of one $SU(2)_A$ instanton $\widetilde{\CM}_{1, SU(2)_A}$, meeting at a point. As an algebraic variety it is generated by operators transforming in the reducible adjoint representation subject to the Joseph relations, which can be extracted from \eref{generalJoseph}. The Hilbert series can again be written using the surgery formula \eref{surgeryformula} as
\be \label{E3}
\begin{split}
H[\widetilde{\CM}_{1, E_3}](t; \vec x,y) &= H[\widetilde{\CM}_{1, SU(3)}](t; \vec x)+H[\widetilde{\CM}_{1, SU(2)_A}](t; y) -1 \\
&=\sum_{m_1=0}^\infty [m_1,m_1]^{SU(3)}_{\vec x} t^{2m_1} + \sum_{m_2=0}^\infty [2m_2]^{SU(2)_A}_{y} t^{2m_2} -1 ~,
\end{split}
\ee
where $\vec x=(x_1, x_2)$ are the fugacities for $SU(3)$ and $y$ is the fugacity for $SU(2)_A$.

The $SU(3)$ factor of the enhanced global symmetry $E_3$ is broken to $SU(2)_B \times U(1)_I$ when one flows away from the fixed point. The $U(1)$ factor is identified with the topological symmetry $U(1)_I$, up to a normalisation of charges that is explained below. The $SU(2)_B$ factor instead combines with the $SU(2)_A$ factor in $E_3$, which acts as a spectator for the breaking, and together they form a global symmetry $SO(4)$. Hence, we decompose the representations of $SU(3)$ in \eref{E3}, whilst keeping the representations of $SU(2)_A$, i.e we break:
\bea
SU(3) \times SU(2)_A \supset SU(2)_B \times SU(2)_A \times U(1)_I \cong SO(4) \times U(1)_I
\eea

A possible projection matrix that maps the weights of $SU(3)$ to $SU(2)_B \times U(1)$ is given by
\bea 
P_{SU(3) \rightarrow SU(2)_B \times U(1)} = \begin{pmatrix} 0 & 1 \\ 2 & 1 \end{pmatrix}~,
\eea
Let $\vec x= (x_1,x_2)$ be the fugacities of $SU(3)$; $z$ and $w$ be those of $SU(2)_B$ and $U(1)$ respectively (the fugacity $w$ for the $U(1)$ factor will be related to the fugacity $q$ for $U(1)_I$ shortly). 
Under the action of this matrix, the weights of the fundamental representation of $SU(3)$ are mapped as follows:
\bea
(1,0) \rightarrow (0,2)~, \qquad (-1,1) \rightarrow (1,-1)~.
\eea
In other words, we have
\bea
x_1 = w^2~,\quad x_2 x_1^{-1}  = z w^{-1}  \qquad \Leftrightarrow \qquad x_1= w^2~, \quad x_2 = z w~.
\eea
The character of the fundamental representation of $SU(3)$ is mapped to that of $SU(2)_B \times U(1)$ as 
\bea
[1,0] = x_1 + x_2 x_1^{-1} + x_2^{-1} = w^2 + z w^{-1} + z^{-1} w^{-1} = [0_2]+[1_{-1}]~,
%[0;2] +[1;-1]~.
\eea
while the adjoint representation decomposes as
\bea
[1,1] \rightarrow [0_0] + [2_0] + [1_3] + [1_{-3}] ~.
\eea
The $U(1)$ charge is a multiple of $3$ for states in the root lattice. To obtain integer instanton numbers $I \in \bZ$, we set $w^3=q$, where $q$ is the fugacity for $U(1)_I$.

Under this map, the Hilbert series of the reduced moduli space of one $SU(3)$ instanton becomes
\bea \label{dec1}
H[\widetilde{\CM}_{1, SU(3)}](t; z, q) =  \sum_{m=0}^\infty \sum_{n_1=0}^m \sum_{n_2=0}^m [n_1+n_2]_z q^{n_1-n_2} t^{2m}~,
\eea
where $z$ is the $SU(2)_B$ fugacity and $q$ is the $U(1)_I$ fugacity.  

The highest weight generating function%
\footnote{The highest weight generating function for group of rank $r$ is defined as follows:
\bea
\cG(t;\mu_i)=\sum_{n_i,k} b_{n_1,...,n_r,k} ~ \mu_1^{n_1}...\mu_r^{n_r}~ t^k
\eea
where $\{\mu_i\}_{i=1}^r$ are highest weight fugacities s.t. $[n_1,...,n_r] \mapsto \mu_1^{n_1}...\mu_r^{n_r}$ and $\{b_{n_1,...,n_r,k} \}$ are the series coefficients.} \cite{Hanany:2014dia} associated to this Hilbert series is 
\bea \label{HWGSU3only}
\cG[\widetilde{\CM}_{1, SU(3)}](t; \mu,q)=\PE \left[ (1+ \mu q +\mu q^{-1} +\mu^2 ) t^2 - \mu^2 t^4 \right]~,
\eea
where $\mu$ is the fugacity for the highest weight of $SU(2)_B$.

Thus, the highest weight generating function for \eref{E3} becomes
\be \label{HWGE3}
\begin{split}
\cG[\widetilde{\CM}_{1, E_3}](t; \mu, \nu, q)  &= \PE \left[ (1+ \mu q+\mu q^{-1} +\mu^2 ) t^2 - \mu^2 t^4 \right] \\
& \quad + \PE[ \nu^2 t^2] -1~,
\end{split}
\ee
where $\mu$ and $\nu$ are the fugacities corresponding to the highest weights of $SO(4) \cong SU(2)_A \times SU(2)_B$. 

The highest weight generating function  \eref{HWGE3} provides five dominant representations that generate the highest weight lattice in a simple way. The information can be read as follows.
Inside the first PE we can identify the $SU(2)_R$ spin $2$ generators: the singlet $S$, the instanton operator $\mu q$ which we denote by $I\equiv I_{1}$, the anti-instanton operator $\mu q^{-1}$ which we denote by $\widetilde{I}\equiv I_{-1}$, and the meson transforming in the adjoint of $SU(2)_B$, $\mu^2$, which we denote by $T^{\alpha \beta}$ and is subject to the traceless condition $T^{\alpha \beta} \epsilon_{\alpha \beta}=0$. We also identify a relation quadratic in the generators and transforming in the adjoint representation of $SU(2)_B$, the term  $- \mu^2 t^4$.
The second PE is the contribution from the spectator $SU(2)_A$, with the only representation $\nu^2$, the inert meson that we denote by $\widetilde{T}^{\dot{\alpha} \dot{\beta}}$.

Eq. \eref{HWGE3} is an expression that carries information about the representation theory more concisely than the Hilbert series and furthermore the lattice it encodes is a complete intersection. However in order to write the relations between the operators on the chiral ring explicitly, we consider what the Joseph relations for $\widetilde{\CM}_{1, E_3}$ imply.

\subsubsection{The generators and their relations}
For the $\widetilde{\CM}_{1, E_3}$ case, the generators are ${\Phi^i}_j$, with $i=1,2,3$ and ${\Phi^i}_i=0$, transforming in the $[1,1;0]$ of $SU(3) \times SU(2)_A$, and $\widetilde{T}^{\dot{\alpha} \dot{\beta}}$ with $\widetilde{T}^{\dot{\alpha} \dot{\beta}} \epsilon_{\dot{\alpha} \dot{\beta}}=0$, transforming in the $[0,0;2]$ of $SU(3) \times SU(2)_A$. The relations can be read off from \eref{generalJoseph}:
\be
\begin{split}
\Sym^2([1,1;0]+[0,0;2])= & \quad \Sym^2[1,1;0]+\Sym^2[0,0;2]+[1,1;2] \quad \text{where}  \\
\Sym^2([1,1;0])= & \quad [2,2;0]+[1,1;0]+[0,0;0] \\
\Sym^2([0,0;2])= & \quad [0,0;4]+[0,0;0]
\end{split}
\ee
Hence the generator ${\Phi^i}_j$ obeys a quadratic relation transforming in the reducible representation $[1,1;0]+[0,0;0]$ whilst $\widetilde{T}^{\dot{\alpha} \dot{\beta}}$ obeys a singlet relation. This is to be expected, since the minimal nilpotent orbit of traceless $2 \times 2$ matrix is the subset of matrices with zero determinant. There is also a quadratic relation mixing ${\Phi^i}_j$ and $\widetilde{T}^{\dot{\alpha} \dot{\beta}}$ transforming in the $[1,1;2]$. We can write these relations as follows:%
\footnote{For $\widetilde{T}$ a symmetric $2 \times 2$ matrix, \ie~ $\tilde{T}^{\dot\alpha \dot\beta} \epsilon_{\dot\alpha \dot \beta} =0$, the following statements are equivalent: $\widetilde{T}^2=0$, $\det \widetilde{T}=0$ and $\Tr \widetilde{T}^2=0$. }
\bea \label{SU3xSU2rel}
\begin{array}{ll}
~[1,1;0]+[0,0;0]& :  \quad {\Phi^i}_j {\Phi^j}_k=0  \\
~[0,0;0] & : \quad \Tr(\widetilde{T}^2) \equiv \widetilde{T}^{\dot{\alpha} \dot{\beta}} \epsilon_{\dot{\alpha} \dot{\sigma}} \epsilon_{\dot{\beta} \dot{\rho}} \widetilde{T}^{\dot{\rho} \dot{\sigma}}= 0 \\
~[1,1;2] &  : \quad {\Phi^i}_j \widetilde{T}^{\dot{\alpha} \dot{\beta}} = 0~,
\end{array}
\eea
where the indices of $\tilde{T}$ are contracted by the epsilon tensor, \eg~ $(\tilde{T}^2)^{\dot \alpha \dot \sigma} = \widetilde{T}^{\dot{\alpha} \dot{\beta}}  \epsilon_{\dot{\beta} \dot{\rho}} \widetilde{T}^{\dot{\rho} \dot{\sigma}}$~.

The glueball operator, the instanton and anti-instanton operators and the meson are embedded into the generator ${\Phi^i}_j$ since this is the one transforming nontrivially under the $SU(3)$ factor that breaks into $SU(2)_B \times U(1)$. We choose the following embedding:
\bea
\begin{array}{lr}
{\Phi^\alpha}_\beta = T^{\alpha\gamma}\epsilon_{\gamma\beta} - \frac{1}{2} S \delta^{\alpha}{}_{\beta}
 &\qquad \alpha,\beta=1,2  \\
{\Phi^\alpha}_3 =I^\alpha \\
{\Phi^3}_\alpha = \epsilon_{\alpha\beta} \tilde{I}^\beta \\
{\Phi^3}_3 =S
\end{array}
\eea
where $T^{\alpha\beta}$ is a traceless $2 \times 2$ matrix, $T^{\alpha \beta} \epsilon_{\alpha \beta}=0$. Notice that the choice of ${\Phi^\alpha}_\beta$ ensures that ${\Phi^i}_j$ is traceless since ${\Phi^i}_i={\Phi^\alpha}_\alpha+{\Phi^3}_3=0$. % The minus sign in ${\Phi^3}_\alpha$ is chosen for later convenience. 

The aim is to decompose the relations in the first and third equations of \eref{SU3xSU2rel}.  Under $SU(3) \times SU(2)_A \supset SU(2)_B \times U(1)_I \times SU(2)_A$ the representations decompose as
\be \label{Adj+SinglDynkin}
\begin{split}
[1,1;0] +[0,0;0] &\quad \rightarrow \quad [2_{0};0]+[1_{1},0]+[1_{-1},0]+2[0_0,0] \\
[1,1;2] &\quad  \rightarrow \quad [2_{0};2]+[1_{1};2]+[1_{-1};2]+[0_0;2]~.
\end{split}
\ee
Thus the relations in the first equation of \eref{SU3xSU2rel} decompose into the five relations%, which can explicitly be written as:
\bea \label{RelSU2}
\begin{array}{rll}
~[2_{0};0]: &\quad S T^{\alpha \beta} = - I^\alpha \widetilde{I}^\beta + \frac{1}{2}(I^ \rho \epsilon_{\rho \sigma} \widetilde{I}^\sigma ) \epsilon^{\alpha \beta}  \\
~[1_{1},0]: &\quad I^\beta \epsilon_{\beta\gamma} T^{\gamma \alpha }  = \frac{1}{2} I^\alpha S \\
~[1_{-1},0]: &\quad \widetilde{I}^\beta \epsilon_{\beta\gamma} T^{\gamma \alpha} = -\frac{1}{2} \widetilde{I}^\alpha S \\
~2[0_0,0]: &\quad S^2=\widetilde{I}^\alpha \epsilon_{\alpha \beta} I^\beta = 2 \Tr(T^2)~.
\end{array}
\eea
The relations in the second line of \eref{Adj+SinglDynkin} can be explicitly written as:
\bea \label{crossRelns}
\begin{array}{rll}
~[2_{0};2]: &\quad T^{\alpha\beta} \widetilde{T}^{\dot{\alpha} \dot{\beta}}=0   \\
~[1_{1};2]: &\quad  I^\alpha \widetilde{T}^{\dot{\alpha} \dot{\beta}}=0 \\
~[1_{-1};2]: &\quad \widetilde{I}^\alpha \widetilde{T}^{\dot{\alpha} \dot{\beta}}=0 \\
~[0_0;2]: &\quad S \widetilde{T}^{\dot{\alpha} \dot{\beta}}=0~.
\end{array}
\eea
Recall also from \eref{SU3xSU2rel} that 
\bea
~[0_0;0]: &\quad \Tr(\widetilde{T}^2) =0~. \label{3rdSingRel}
\eea
In total there are thus 10 equations, namely  \eref{RelSU2}, %\eref{2ndSingRel},
 \eref{crossRelns} and \eref{3rdSingRel}.%
\footnote{Notice that the meson $\widetilde{T}^{\dot{\alpha} \dot{\beta}}$, the generator for the spectator $SU(2)_A$, is made up of the same fundamental fields (quarks) as the meson $T^{\alpha \beta}$. Before considering gauge invariant combinations, the quarks ${Q^{ \alpha \dot{\alpha}}}_{\mathfrak{a}}$, with $\alpha, \dot{\alpha}=1,2$ and $\mathfrak{a}=1,2$, transform in the vector representation of the global symmetry $SO(4) \cong SU(2)_A \times SU(2)_B$ and in the fundamental representation of the \emph{gauge} group $SU(2)$. Out of these quarks the following gauge invariant mesons can be constructed: $ T^{\alpha \beta}={Q^{\alpha \dot{\alpha}}}_{\mathfrak{a}} {Q^{\beta \dot{\beta}}}_{\mathfrak{b}} \epsilon^{\mathfrak{a} \mathfrak{b}} \epsilon_{\dot{\alpha} \dot{\beta}}$ and $\widetilde{T}^{\dot{\alpha} \dot{\beta}}= {Q^{\alpha \dot{\alpha}}}_{\mathfrak{a}} {Q^{\beta \dot{\beta}}}_{\mathfrak{b}} \epsilon^{\mathfrak{a} \mathfrak{b}}\epsilon^{\alpha \beta} $. The difference between these two mesons is in the relations they satisfy at infinite coupling, one being quantum corrected whilst the other being unaffected: $\Tr(\widetilde{T}^2)=0$ vs
$2\Tr(T^2)= S^2 = I \cdot \widetilde{I}$.}

The finite coupling result that $S$ be nilpotent is obtained by virtue of the last equation of \eref{RelSU2} when we set $I, \widetilde{I} =0$. Consequently we also restore the condition $\Tr(T^2)=0$, which, for a traceless $2 \times 2$ matrix, is equivalent to $T^2=0$, the classical relation. Moreover \eref{SMreln} is also recovered.

Another approach to see these 10 relations between the operators at infinite coupling is to rewrite \eref{E3} in terms of characters of representations of $SO(4) \times U(1)$ and compute its plethystic logarithm.  For reference, we present such a Hilbert series up to order $t^4$ as follows: 
\bea
&H[E_3](t; x_1, x_2, q) =  1+ \Big(1 + [2,0]+[0,2] + (q +q^{-1})[1,0] \Big) t^2 +  \label{HS1Nf2}\\
&+ \Big(1 + [2,0] + [4,0] +[0,4] + (q+q^{-1})([1,0]+[3,0]) +(q^2+q^{-2}) [2,0] \Big) t^4 + \ldots~.\nonumber
\eea
The plethystic logarithm of this Hilbert series is
\be\label{PL1Nf2}
\begin{split}
\PL &\left[ H[E_3](t; x_1, x_2, q) \right] = \Big(1 + [2,0]+[0,2]+ (q +q^{-1})[1,0] \Big) t^2  +\\
& \quad - \Big(3+ [2,0]+[0,2]+ [2,2] + (q+q^{-1}) ([1,2]+[1,0]) \Big)t^4 + \ldots~.
\end{split}
\ee
Indeed, the 10 relations listed in \eref{RelSU2}, \eref{crossRelns} and \eref{3rdSingRel} are in correspondence with the terms at order $t^4$ in \eref{PL1Nf2}.  We emphasise here that the computation of the plethystic logarithm provides an efficient way to write down the relations that are crucial to describe the moduli space.  This method is applied for the cases of higher $N_f$ in subsequent sections.

We can rewrite these relations in terms of a $4 \times 4$ adjoint matrix $M^{ab}$, with $a,b,c,d=1,\ldots, 4$ vector indices of $SO(4)$, such that
\bea
M^{ab} = - M^{ba}~,
\eea
as follows:
\bea 
~[2,2]+[0,0]: &\quad M^{ab}M^{bc} =  (\epsilon_{\alpha \beta} I^\alpha \widetilde{I}^\beta) \delta^{ac} \label {FirstExplicitReln2Flav} \\
~[0,0]: &\quad \epsilon_{abcd} M^{ab}M^{cd} = \epsilon_{\alpha \beta}  I^\alpha \widetilde{I}^\beta \\
~[0,0]: &\quad S^2 = \epsilon_{\alpha \beta} I^\alpha \widetilde{I}^\beta \label{gauginoInstanReln2Flav}\\
~[2,0]: &\quad S M^{ab} (\gamma^{ab})^{\alpha \beta} =\widetilde{I}^{(\alpha} I^{\beta)}  \\
~[0,2]: &\quad S M^{ab} (\gamma^{ab})_{\dot \alpha \dot \beta} =0  \label{orthoSandTdotted}\\
~q([1,2]+[1,0]): &\quad  M^{ab} I^\beta (\gamma^{b})_{\beta \dot \alpha}  = SI^\beta (\gamma^a)_{\beta \dot \alpha} \\
~q^{-1}([1,2]+[1,0]): &\quad  M^{ab} \widetilde{I}^\beta (\gamma^b)_{\beta \dot \alpha}  =S \widetilde{I}^\beta (\gamma^a)_{\beta \dot \alpha}\label {LastExplicitReln2Flav} ~.
\eea
The gamma matrices $\gamma^a$ for $SO(4)$ take the following index form:
\bea
(\gamma^a)_{\alpha \dot \alpha}~
\eea
and the product of two gamma matrices is defined as:
\bea
(\gamma^{ab})_{\alpha \beta} \equiv (\gamma^{[a})_{\alpha \dot \alpha} (\gamma^{b]})_{ \beta \dot \beta} \epsilon^{\dot \alpha \dot \beta} \quad \text{and} \quad (\gamma^{ab})_{\dot \alpha \dot \beta} \equiv (\gamma^{[a})_{\alpha  \dot\alpha } (\gamma^{b]})_{ \beta \dot \beta} \epsilon^{\alpha \beta}~;
\eea
where the spinor indices are raised and lowered with the epsilon tensor.

\subsubsection{Expansion in the instanton fugacity}

It is useful to rewrite \eref{HWGE3} in terms of an expansion in $q$:
\bea \label{qexpNf2}
\cG[\widetilde{\CM}_{1, E_3}](t; \mu, \nu, q) &=\frac{1}{(1-t^2)(1-t^2 \mu^2)} \sum^{\infty}_{n = -\infty } q^{n}  t^{2 |n|} \mu^{|n|}+\frac{1}{1-\nu^2 t^2}-1
\eea
From here, we can extract the transformation properties of instanton operators of charge $n$ under the $U(1)_I$. They transform as spin $|n|$ highest weight states for $SU(2)_R$ and as spin $|n|/2$ representations of $SU(2)_B$.

The classical dressing for each $q^n$ instanton operator, the factor outside the sum, is, for $n \neq 0$, a space generated by the $SU(2)_B$ adjoint meson $T^{\alpha \beta}=M^{ab}(\gamma^{ab})^{\alpha \beta}$ and the glueball operator $S$ obeying the relation:
\bea \label{SO4dressingNonzeroSect}
\Tr(T^2)=S^2
\eea

For $n=0$ there is a contribution coming from the $SU(2)_A$, the second term in \eref{qexpNf2}, which modifies the classical dressing entirely.  The latter is in fact, for this charge zero sector, generated by $M^{ab}$ and $S$ subject to the following relations:
\bea \label{SO4dressingZeroSect}
~[2,2]+[0,0]: & \quad M^{ab} M^{bc}=S^2 \delta^{ac} \\
~[0,0]: & \quad \epsilon_{abcd} M^{ab} M^{cd}=S^2 \\
~[0,2]: &\quad S M^{ab} (\gamma^{ab})_{\dot \alpha \dot \beta} =0 
\eea
These relations are a subset of \eref{FirstExplicitReln2Flav} - \eref{LastExplicitReln2Flav} constructed in the following way:  we take the first two equations and  we substitute the instanton bilinear on the right hand side with the glueball operator by means of \eref{gauginoInstanReln2Flav}. Moreover we keep \eref{orthoSandTdotted} as it is a relation not corrected by instanton operators.

\subsection{$N_f =3$}

The moduli space of the reduced one $E_4=SU(5)$ instanton, $\widetilde{\CM}_{1, E_4= SU(5)}$, is the nilpotent orbit generated by the adjoint representation of $SU(5)$. Its associated Hilbert series can thus be written as
\bea \label{HSSU5}
H[\widetilde{\CM}_{1, SU(5)}](t; \vec x) = \sum_{n=0}^\infty [n,0,0,n]_{\vec x} t^{2n}~,
\eea
where $[1,0,0,1]_{\vec{x}}$ is the character of the adjoint representation of $SU(5)$ with fugacities $\vec{x}=(x_1,x_2,x_3,x_4)$.
In order to proceed with a decomposition from weights of $SU(5)$ representations to those of $SO(6) \times U(1)$, we choose the projection matrix
\bea
P_{A_4 \rightarrow D_3\times U(1)} = \left(
\begin{array}{cccc}
 0 & 0 & 1 & 0 \\
 0 & 0 & 0 & 1 \\
 0 & 1 & 0 & 0 \\
 4 & 3 & 2 & 1 \\
\end{array}
\right)~,
\eea
which gives the fugacity map 
\be
\begin{split}
&x_1 = w^4~, \quad x_2 x_1^{-1} = y_3 w^{-1}~, \quad x_3 x_2^{-1} = y_1 y_3^{-1} w^{-1}~, \quad x_4 x_3^{-1} = y_1^{-1} y_2 w^{-1}~,  \\
& \Leftrightarrow \quad x_1 = w^4~, \qquad x_2  = y_3 w^{3}~, \qquad x_3 = y_1 w^{2}~, \qquad x_4 = y_2 w~.
\end{split}
\ee
States in the root lattice carry a charge multiple of $5$ for the $U(1)$ associated to the fugacity $w$, hence we set $w^5=q$ in the following, where $q$ is the fugacity for the integer quantized instanton number $U(1)_I$.
Then \eref{HSSU5} can be written in terms of the character expansion of $SO(6) \times U(1) \supset  SU(5)$ as
\bea \label{SU5dec1}
H[\widetilde{\CM}_{1, SU(5)}](t; \vec y, q) = \sum_{n=0}^\infty \sum_{n_1=0}^n \sum_{n_2=0}^n [0, n_1, n_2]_{\vec y} q^{n_1 - n_2} t^{2n} ~,
\eea
where $[p_1,p_2,p_3]_{\vec{y}}$ is the character of a representation of $SO(6)$ as a function of fugacities $\vec{y}=(y_1,y_2,y_3)$.
The information contained in this equation can be carried compactly by means of the associated highest weight generating function
\bea \label{HWGNf3}
\cG[\widetilde{\CM}_{1, SU(5)}](t; \mu_2, \mu_3; q) = \PE \left[ t^2(1+\mu_2 q +\mu_3 q^{-1}+\mu_2 \mu_3) -t^4 \mu_2 \mu_3 \right]~
\eea
where at $t^2$ we can again recognise the contribution of $S$, a singlet of $SO(6)$, the instanton and the anti-instanton operators in the spinor $[0,1,0]$ and cospinor $[0,0,1]$ representations, and the meson in the adjoint representation $[0,1,1]$, while at order $t^4$ is the basic relation between the operators.
Notice that \eref{HWGNf3} is a generating function for a lattice with conifold structure.

\subsubsection{The generators and their relations}

The generators and the relations can be extracted from the plethystic logarithm of the Hilbert series. The Hilbert series of the reduced moduli space of $1$ $E_4$ instanton can be written in terms of characters of $SO(6) \times U(1)$ up to $O(t^4)$ as:
\be\label{HSE4Nf3}
\begin{split}
H[E_4](t; \vec x, q) &= 1 +  (1 + [0, 1, 1] + q^{-1} [0, 0, 1] + q [0, 1, 0])t^2 +   \\
& + \Big(1 + [0, 1, 1] + [0, 2, 2] +q^{-1} ( [0, 0, 1] + [0, 1, 2] )  +\\
&+ q ([0,1,0] + [0,2,1] )+ q^{-2} [0, 0, 2] + q^2 [0, 2, 0] \Big) t^4 + \ldots~. 
\end{split}
\ee
The plethystic logarithm of this Hilbert series is
\be\label{HE4plSO6}
\begin{split}
& \PL \left[ H[E_4](t; \vec x, q) \right] \\
&=  (1 + [0, 1, 1]+ q^{-1} [0, 0, 1] +  q [0, 1, 0] ) t^2 - \Big(2 + 2 [0, 1, 1]+ [2, 0, 0]  +\\
&+ q ([1, 0, 1] + [0, 1, 0]) + q^{-1} ([1, 1, 0] + [0, 0, 1]) \Big)  t^4 + \ldots~. 
\end{split}
\ee
Below we write down the generators corresponding to the terms at $t^2$ and the explicit relations corresponding to the terms at order $t^4$ of \eref{HE4plSO6}. 

For $SO(6)$, we use $a,b,c,d =1, \ldots, 6$ to denote vector indices and use $\alpha, \beta, \rho, \sigma=1,\ldots, 4$ to denote spinor indices.  Note that the spinor representation of $SO(6)$ is complex. The delta symbol carries has one upper and one lower index:
\bea
\delta^{\alpha}_{\beta}~.
\eea
The gamma matrices $\gamma^a$ can take the following forms:
\bea
(\gamma^a)_{\alpha \beta}  \qquad \text{and} \qquad (\gamma^b)^{\alpha \beta}~,
\eea
where the $\alpha, \beta$ indices are antisymmetric.  The product of two gamma matrices has one lower spinor index and one upper spinor index:
\bea
(\gamma^{ab})^{\alpha}_{~\rho} \equiv (\gamma^{[a})^{\alpha \beta} (\gamma^{b]})_{\beta \rho}~.
\eea
From \eref{HE4plSO6} the generators of the moduli space are $M^{ab}$, a $6 \times6$ antisymmetric matrix, the instanton operators $I^\alpha$ and $\widetilde{I}_{\alpha}$ and the gaugino bilinear $S$. The relations corresponding to the terms at order $t^4$ of \eref{HE4plSO6} can be written as follows:
\bea 
[2, 0, 0]+[0,0,0]: & \qquad  M^{ab}M^{bc}= (I^\alpha \widetilde{I}_{\alpha} ) \delta^{ac} \label{FirstExplicitReln3Flav}\\
[0,1,1]: & \qquad \epsilon^{abcdef} M^{cd} M^{ef} = \widetilde{I}_\beta (\gamma^{ab})^\beta_{~\alpha}  I^\alpha \\
[0,0,0]:& \qquad S^2 = I^\alpha \widetilde{I}_{\alpha}  \\
[0,1,1]: & \qquad S M^{ab} = \widetilde{I}_\beta (\gamma^{ab})^\beta_{~\alpha}  I^\alpha  \\
q([1,0,1]+[0,1,0]): & \qquad M^{ab} I^\alpha (\gamma^{b})_{\alpha \beta}  = S I^{\alpha} (\gamma^a)_{\alpha \beta}  \\
q^{-1}([1,1,0]+[0,0,1]): & \qquad M^{ab} \widetilde{I}_\alpha (\gamma^{b})^{\alpha \beta} = S \widetilde{I}_\alpha (\gamma^a)^{\alpha \beta} \label{LastExplicitReln3Flav}~.
\eea
As can be seen, the classical relations are corrected by instanton bilinears and this is a recurrent feature for all number of flavours.
These relations can also be rewritten in terms of an $SU(4)$ matrix $M^{\alpha}_{~\beta}$ using the following relation
\bea
M^{ab} = M^{\alpha}_{~\beta} (\gamma^{ab})^{\beta}{}_{\alpha}~.
\eea

\subsubsection{Expansion in the instanton fugacity }

We rewrite \eref{HWGNf3} as an expansion in $q$ as follows:
\bea \label{qexpNf4}
\cG[\widetilde{\CM}_{1, SU(5)}](t; \mu_2, \mu_3, q) &= \frac{1}{(1-t^2)(1-t^2 \mu_2\mu_3)}\Big(\sum_{n \geq 0} q^{n}  (t^2 \mu_2)^n   + \sum_{n<0} q^{n}  (t^2 \mu_3)^{-n} \Big)~.
\eea

Two very interesting features emerge from the $q$ expansion. Firstly, an instanton operator of charge $n$ has $SU(2)_R$ spin $|n|$ and it transforms as an $|n|$-spinor --- a representation with $|n|$ on a spinor Dynkin label --- of the global flavour group $SO(6)$.  Whilst in \cite{Tachikawa:2015mha} it was found that this result holds for $n=1$, here we find a prediction for all $n$.  

Secondly the instanton operators are dressed by a factor, the one in front of the sum, which is generated by $S$ and $M^{ab}$, subject to the following relations: 
\bea
[2,0,0]+[0,0,0]: & \qquad M^{ab} M^{bc} =S^2 \delta^{ac}  \\
[0,1,1]: & \qquad \epsilon^{abcdef} M^{cd} M^{ef} =S M^{ab}~. 
\eea
Interestingly, such relations can be extracted directly from \eref{FirstExplicitReln3Flav} - \eref{LastExplicitReln3Flav} by keeping only those relations that are not corrected by the instanton operators. This feature is a recurrent theme for higher number of flavours.

\subsection{$N_f =4$}

The Higgs branch at infinite coupling for an $SU(2)$ theory with $N_f=4$ flavours is isomorphic to the reduced moduli space of one $E_5=SO(10)$ instanton $\widetilde{\CM}_{1, E_5= SO(10)}$, which is given by the minimal nilpotent orbit of $SO(10)$. Its Hilbert series is
\bea
H[\widetilde{\CM}_{1, SO(10)}](t; \vec x) = \sum_{n=0}^\infty [0,n,0,0,0]_{\vec x} t^{2n}~,
\eea
where $[0,1,0,0,0]_{\vec x} $ is the character of the adjoint representation of $SO(10)$.

At finite coupling the theory has a global symmetry $SO(8) \times U(1)$. Hence we rewrite this Hilbert series in terms of an $SO(8) \times U(1)$ character expansion as
\bea \label{SO10dec1}
H[\widetilde{\CM}_{1, SO(10)}](t; \vec y, q) = \frac{1}{1-t^2} \sum_{n_1, n_2, n_3 \geq 0} [0, n_1, 0, n_2+n_3]_{\vec y} q^{n_2 - n_3} t^{2n_1+2n_2+2n_3} ~,
\eea
where we decompose representations of $SO(8) \times U(1) \subset SO(10)$ using a projection matrix that maps the weights of $SO(10)$ representations to those of $SO(8) \times U(1)$ as follows
\bea
P_{D_5 \rightarrow D_4\times U(1)} = \left(
\begin{array}{ccccc}
 0 & 0 & 0 & 0 & 1 \\
 0 & 0 & 1 & 0 & 0 \\
 0 & 0 & 0 & 1 & 0 \\
 0 & 1 & 0 & 0 & 0 \\
 -2 & -2 & -2 & -1 & -1 \\
\end{array}
\right)~.
\eea
Under the action of this matrix, the fugacities $\vec x$ of $SO(10)$ are mapped to the fugacities $\vec y$ of $SO(8)$ and $w$ of $U(1)$ as follows:
\be
\begin{split}
 (x_1, x_2 x_1^{-1} , x_3 x_2^{-1},  x_4 x_5 x_3^{-1}, x_5 x_4^{-1}) &=  \left( w^{-2},  y_4, y_2 y_4^{-1} , y_1 y_2^{-1} y_3, y_1 y_3^{-1}  \right) \\
 \Leftrightarrow \qquad (x_1, x_2,x_3,x_4,x_5) &= \left( \frac{1}{w^2},\frac{y_4}{w^2},\frac{y_2}{w^2},\frac{y_3}{w},\frac{y_1}{w} \right)~.
\end{split}
\ee
In \eref{SO10dec1} we set $w^2=q$ to have integer instanton numbers, rather than even.

The corresponding highest weight generating function is
\bea \label{E5HWG}
\cG[\widetilde{\CM}_{1, SO(10)}] (t; \mu_2, \mu_4 ; q)= \PE \left[ t^2(1+ \mu_2 +\mu_4 q + \mu_4 q^{-1}) \right]
\eea
where we recognise the usual $SU(2)_R$ spin-$2$ generators: the glueball superfield $S$, a singlet of $SO(8)$, the instanton operators $I_\alpha$ and $\widetilde{I}_\alpha$ associated to $\mu_4 q $ and $\mu_4 q^{-1} $, both transforming in the same spinor representation of $SO(8)$ with opposite $U(1)$ charge, as well as the meson $M^{ab}$, associated to $\mu_2$. The highest weight  lattice is freely, generated as we see from the lack of relations at order $t^4$.

\subsubsection{The generators and their relations}
The expansion of \eref{SO10dec1} up to order $t^4$ is given by
\be
\begin{split}
H[E_5]&(t; \vec x, q)= 1 + \Big(1 + [0, 1, 0, 0]+ (q+q^{-1}) [0, 0, 0, 1] \Big) t^2 +  \\
&+\Big(1+ [0, 1, 0, 0]  + [0, 0, 0, 2]  + [0, 2, 0, 0]+ \\
&+(q+q^{-1})([0, 0, 0, 1]+[0, 1, 0, 1]) + (q^2+q^{-2}) [0, 0, 0, 2]   \Big) t^4+\ldots~.
\end{split}
\ee
The plethystic logarithm of this Hilbert series is
\be \label{plE5SO8}
\begin{split}
 \PL &\left[ H[E_5](t; \vec x, q) \right] =\Big(1+ [0, 1, 0, 0] + (q+q^{-1}) [0, 0, 0, 1] \Big) t^2 +\\
&\qquad - \Big( 2+[2,0,0,0]+[0,1,0,0]+[0,0,2,0]+[0,0,0,2]  +\\
&\qquad +(q+q^{-1})( [1, 0, 1, 0] +[0,0,0,1]) + (q^2+q^{-2})\Big) t^4+\ldots~.
\end{split}
\ee
From this collection of representations we can write the defining equations for the Higgs branch at infinite coupling by constructing the relevant operators. For $SO(8)$, we use $a,b,c,d =1, \ldots, 8$ to denote the vector indices, $\alpha, \beta, \rho, \sigma=1,\ldots, 8$ to denote those in the spinor representation $[0,0,0,1]$ and $\dot \alpha, \dot \beta, \dot \rho, \dot \sigma=1,\ldots, 8$ to denote those in the conjugate spinor representation $[0,0,1,0]$.  The delta symbol has the following forms:
\bea
\delta^{\alpha \beta}~ \quad \text{or}~\quad \delta_{\alpha \beta}~ \quad \text{or}~\quad \delta^{\dot \alpha \dot \beta}~ \quad \text{or}~\quad \delta_{\dot \alpha \dot \beta}~.
\eea
The gamma matrices $\gamma^a$ can take the following forms:
\bea
(\gamma^a)_{\alpha \dot \alpha}  \qquad \text{or} \qquad (\gamma^a)^{\alpha \dot \alpha}~.
\eea
The product of two gamma matrices has the following forms:
\be
\begin{split}
(\gamma^{ab})_{\alpha \beta} \equiv (\gamma^{[a})_{\alpha \dot \beta} (\gamma^{b]})_{\beta \dot \beta}  \quad \text{and} \quad (\gamma^{ab})_{\dot \alpha \dot \beta} \equiv (\gamma^{[a})_{\alpha \dot \alpha} (\gamma^{b]})_{\alpha \dot \beta} 
\end{split}
\ee
and similarly for both upper indices; the indices $\alpha, \beta$ and $\dot \alpha, \dot \beta$ are antisymmetric.  The product of four gamma matrices has the following forms:
\be
\begin{split}
(\gamma^{abcd})_{\alpha \beta} &\equiv (\gamma^{[a})_{\alpha \dot \beta} (\gamma^b)_{\rho \dot \beta} (\gamma^c)_{\rho \dot \sigma}  (\gamma^{d]})_{\beta \dot \sigma}  \\ %\quad \text{and}  \nn \\
(\gamma^{abcd})_{\dot \alpha \dot \beta} &\equiv (\gamma^{[a})_{\alpha \dot \alpha} (\gamma^b)_{\alpha \dot \rho} (\gamma^c)_{\rho \dot \rho}  (\gamma^{d]})_{\rho \dot \beta}  
\end{split}
\ee
and similarly for both upper indices; the indices $\alpha, \beta$ and $\dot \alpha, \dot \beta$ are symmetric. 

The generators of the moduli space are $M^{ab}$, which is a $8 \times8$ antisymmetric matrix;  the instanton operators $I_{\alpha}$ and $\widetilde{I}_{\alpha}$; and the glueball superfield $S$.

The relations corresponding to terms at order $t^4$ of \eref{plE5SO8} can be written as
\bea
[2,0,0,0]+[0,0,0,0]: & \qquad M^{ab} M^{bc} =(I_{\alpha} \widetilde{I}_{\alpha}) \delta^{ac}  \label{FirstRelns4flav} \\
[0,0,2,0]: & \qquad M^{ab} M^{cd} (\gamma^{abcd})_{\dot \alpha \dot \beta} =0 \\
[0,0,0,2]: & \qquad M^{ab} M^{cd} (\gamma^{abcd})_{\alpha \beta} =  I_{(\alpha} \widetilde{I}_{\beta)} - \frac{1}{8} (I_\rho \widetilde{I}_\rho) \delta_{\alpha \beta}  \\
[0,0,0,0]: &\qquad S^2 = I_{\alpha} \widetilde{I}_{\beta} \delta^{\alpha \beta} \\
[0,1,0,0]: & \qquad S M^{ab} = I_\alpha \widetilde{I}_\beta (\gamma^{ab})_{\alpha \beta} \\
q([1,0,1,0]+[0,0,0,1]): &\qquad M^{ab} I_\beta (\gamma^b)_{ \beta \dot \alpha} =  S I_\beta (\gamma^a)_{\beta \dot \alpha}\\
q^{-1}([1,0,1,0]+[0,0,0,1]): &\qquad M^{ab} \widetilde{I}_\beta (\gamma^b)_{ \beta \dot \alpha} = S \widetilde{I}_\beta (\gamma^a)_{\beta \dot \alpha} \\
(q^2+q^{-2})[0,0,0,0]:  &\qquad I_\alpha I_\beta \delta_{\alpha \beta} =  \widetilde{I}_\alpha \widetilde{I}_\beta \delta_{\alpha \beta} =0~  \label{LastRelns4flav}.
\eea

\subsubsection{Expansion in the instanton fugacity }

In terms of an expansion in $q$, \eref{E5HWG} can be written as
\bea
\cG[\widetilde{\CM}_{1, SO(10)}](t; \mu_2, \mu_4 ; q)&= \frac{1}{(1-t^2)(1-\mu_2 t^2)(1-\mu_4^2 t^4)}\sum_{n =-\infty}^\infty q^{n} \mu_4^{|n|} t^{2|n|} ~.
\eea
Here again we find that instanton operators of charge $n$ are spin $|n|$ of $SU(2)_R$ and transform in $|n|$-spinor representations of $SO(8)$. 

However the interpretation of the classical dressing is more subtle than in previous cases. The prefactor in the $q$ expansion signifies a space which is algebraically determined by some of the conditions that define the moduli space of one $SO(8)$ instanton; in particular it is a space generated by two operators, $M^{ab}$, in the adjoint representation $[0,1,0,0]$ of $SO(8)$, and $S$, in the singlet $[0,0,0,0]$, subject to relations that transform in the representations $[2,0,0,0]$, $[0,0,0,0]$ and $[0,0,2,0]$. Explicitly these relations are:
\bea
[2,0,0,0]+[0,0,0,0]: & \qquad M^{ab} M^{bc} =S^2 \delta^{ac}  \\
[0,0,2,0]: & \qquad M^{ab} M^{cd} (\gamma^{abcd})_{\dot \alpha \dot \beta} =0~. 
\eea
The following features can be observed. Whilst the classical moduli space of one $SO(8)$ instanton is generated by \eref{ExplicitJosephForSO2N-Symm} and \eref{ExplicitJosephForSO2N-Antisymm}, here the anti-self-dual 4th rank antisymmetric representation is missing%
\footnote{Recall that for $SO(8)$, $\wedge^4 [1,0,0,0]=[0,0,2,0]+[0,0,0,2]$ is a reducible representation.}.
Such a space has complex dimension 13 and, by adding the dimension originating from the sum over the instanton number, the correct 14 dimensional moduli space of one $SO(10)$ instanton is recovered.
Again, the classical dressing can be guessed from the set of equations in \eref{FirstRelns4flav}-\eref{LastRelns4flav} by keeping only the relations that are not corrected by the instanton operators.

\subsection{$N_f=5$}

The Hilbert series of $\widetilde{\CM}_{1, E_6}$ can be written as
\bea
H[\widetilde{\CM}_{1, E_6}](t; \vec x) = \sum_{n=0}^\infty {\Esix0  n 0 0 0 0}_{\vec x} t^{2n}~.
\eea
A projection matrix that maps the weights of $E_6$ to those of $D_5 \times U(1)$ is given by
\bea
P_{E_6 \rightarrow D_5 \times U(1)} =\left(
\begin{array}{cccccc}
 0 & 0 & 0 & 0 & 0 & 1 \\
 0 & 0 & 0 & 0 & 1 & 0 \\
 0 & 0 & 0 & 1 & 0 & 0 \\
 0 & 0 & 1 & 0 & 0 & 0 \\
 0 & 1 & 0 & 0 & 0 & 0 \\
 -4 & -3 & -5 & -6 & -4 & -2 \\
\end{array}
\right)~.
\eea
Under the action of this matrix, the fugacities of $\vec x$ of $E_6$ are mapped to the fugacities $\vec y$ of $SO(10)$ and $w$ of $U(1)$ as follows:
\bea
 (x_1, x_1 x_2^{-1} , x_1 x_3^{-1},  x_2 x_6^{-1}, x_3 x_5^{-1}, x_3 x_4^{-1}) &= \left( \frac{1}{w^4}, \frac{1}{w y_5}, \frac{w}{y_4},  \frac{y_5}{w y_1},\frac{y_4}{w y_2}, \frac{w y_4}{y_3}\right) \nn\\
\Leftrightarrow \qquad (x_1,x_2,x_3,x_4,x_5,x_6) &= \left( \frac{1}{w^4},\frac{y_5}{w^3},\frac{y_4}{w^5},\frac{y_3}{w^6},\frac{y_2}{w^4}, \frac{y_1}{w^2}\right)~.
\eea
The fugacity of $U(1)_I$ is $q=w^3$. Thus, the Hilbert series $H[\widetilde{\CM}_{1, E_6}]$ can be written in terms of characters of representations of $SO(10) \times U(1)_I$ as
\bea \label{E6dec1}
H[\widetilde{\CM}_{1, E_6}](t; \vec y, q) = \frac{1}{1-t^2} \sum_{n_1, n_2, n_3 \geq 0} [0, n_1, 0, n_2,n_3]_{\vec y} q^{n_2 - n_3} t^{2n_1+2n_2+2n_3} ~,
\eea
The corresponding highest weight generating function is
\bea \label{E6decomHWG}
\cG[\widetilde{\CM}_{1, E_6}] (t; \mu_2, \mu_4 ; q)= \PE \left[ t^2(1+ \mu_2 +\mu_4 q + \mu_5 q^{-1}) \right]~.
\eea

\subsubsection{The generators and their relations}
The expansion of \eref{E6dec1} up to order $t^4$ is given by
\be\label{HSE6Nf5}
\begin{split}
H[E_6]&(t; \vec x, q) = 1 +  (1 + [0, 1, 0, 0, 0]+ q^{-1}[0, 0, 0, 0, 1]+ q [0, 0, 0, 1, 0])t^2 +\\
&+ \Big(1 + [0, 1, 0, 0, 0]  +[0, 2, 0, 0, 0] + [0, 0, 0, 1, 1] +\\
&\quad + q^{-1} ([0, 0, 0, 0, 1] +[0, 1, 0, 0, 1]) + q ([0, 0, 0, 1, 0] +[0, 1, 0, 1, 0]) +\\
&\quad + q^{-2} [0, 0, 0, 0, 2] + q^2 [0, 0, 0, 2, 0] \Big)t^4  +\ldots~. 
\end{split}
\ee
The plethystic logarithm of this Hilbert series is
\be\label{HE6pla}
\begin{split}
\PL &\left[ H[E_6](t; \vec x, q) \right] =  (1+ [0, 1, 0, 0, 0]+ q^{-1}[0, 0, 0, 0, 1]+ q [0, 0, 0, 1, 0] )t^2 + \\ 
&  - \Big (2+[0,1,0,0,0]+ [2,0,0,0,0]+[0,0,0,1,1]+ \\
& \quad + q ([1,0,0,0,1]+[0,0,0,1,0])+q^{-1} ([1,0,0,1,0]+[0,0,0,0,1])+  \\
& \quad + (q^2+q^{-2})[1,0,0,0,0] \Big )t^4+\ldots~. 
\end{split}
\ee

For $SO(10)$, we use $a,b,c,d =1, \ldots, 10$ to denote vector indices and $\alpha, \beta, \rho, \sigma=1,\ldots, 16$ to denote spinor indices. Note that the spinor representation of $SO(10)$ is complex.  The delta symbol has the following form:
\bea
\delta^\alpha_\beta~.
\eea
The gamma matrices $\gamma^a$ can take the following forms:
\bea
(\gamma^a)_{\alpha  \beta}  \qquad \text{and} \qquad (\gamma^a)^{\alpha \beta}~,
\eea
where the $\alpha, \beta$ indices are symmetric.  The product of two gamma matrices has the following form:
\bea
(\gamma^{ab})^{\alpha}_{~\rho} \equiv (\gamma^{[a})^{\alpha\beta} (\gamma^{b]})_{\beta \rho} ~.
\eea
The product of four gamma matrices has the following form:
\bea
(\gamma^{abcd})^{\alpha}_{~ \beta} &\equiv (\gamma^{[a})^{\alpha\sigma_1} (\gamma^b)_{\sigma_1\sigma_2} (\gamma^c)^{\sigma_2\sigma_3}  (\gamma^{d]})_{\sigma_3 \beta} ~.
\eea

The generators of the moduli space are $M^{ab}$, which is a $10 \times10$ antisymmetric matrix;  the instanton operators $I^{\alpha}$ and $\widetilde{I}_{\alpha}$; and the gaugino superfield $S$.

The relations appearing in the plethystic logarithm \eref{HE6pla} are as follows:
\bea
[2,0,0,0,0]+[0,0,0,0,0]: & \qquad  M^{ab} M^{bc} = ( I^{\alpha} \widetilde{I}_{\alpha}) \delta^{ac}~, \label{FirstExplicitE6relns} \\
[0,0,0,1,1]: & \qquad  M^{[a_1 a_2} M^{a_3 a_4]}= \tilde{I}_\beta  (\gamma^{a_1 \cdots a_{4}})^\beta_{~\alpha}  I^\alpha ~, \\
[0,0,0,0,0]:& \qquad S^2 = I^{\alpha} \widetilde{I}_{\alpha}  ~, \\
[0,1,0,0,0]: &  \qquad  S M^{ab} = \widetilde{I}_{\beta} (\gamma^{ab})^{\beta}_{~\alpha} I^\alpha ~, \\
q([1,0,0,0,1]+ [0,0,0,1,0]): & \qquad M^{ab} I^\alpha (\gamma^b)^\beta_{~\alpha}= S I^\alpha (\gamma^a)^\beta_{~\alpha}~,  \\
q^{-1}([1,0,0,0,1]+ [0,0,0,1,0]): & \qquad M^{ab} \widetilde{I}_\beta (\gamma^b)^\beta_{~\alpha}  = S \widetilde{I}_\beta (\gamma^a)^\beta_{~\alpha}~, \\
(q^2+q^{-2})[1,0,0,0,0]: & \qquad I^\alpha I^\beta (\gamma^a)_{\alpha \beta} = \widetilde{I}_\alpha \widetilde{I}_\beta (\gamma^a)^{\alpha \beta} = 0~ \label{LastExplicitE6relns}.
\eea

\subsubsection{Expansion in the instanton fugacity }

The highest weight generating function \eref{E6decomHWG} can be expanded in the instanton number fugacity $q$ as
\be
\begin{split}
\cG[\widetilde{\CM}_{1, E_6}](t; \mu_2, \mu_4,, \mu_5 ; q) &=\frac{1}{(1-t^2)(1-t^2 \mu_2)(1-t^4 \mu_4\mu_5)} \times \\
& ~~~\quad \Big(\sum_{n \geq 0} q^{n}  (t^2 \mu_4)^n  + \sum_{n<0} q^{n}  (t^2 \mu_5)^{-n}\Big)~.
\end{split}
\ee
From this formula we see that the instanton operators of charge $n$ are spin $|n|$ highest weight states under $SU(2)_R$ and transform in the $n$-spinor representation $[0,0,0,n,0]$ of $SO(10)$ for $n>0$ and the conjugate $|n|$-spinor representation $[0,0,0,0,|n|]$ for $n<0$. 

The dressing factor has the features previously encountered in that is generated by the classical operators $M^{ab}$ and $S$, subject to the relations
\bea
[2,0,0,0,0]+[0,0,0,0,0]: & \qquad  M^{ab} M^{bc} = S^2 \delta^{ac}~.
\eea
Comparing this space to the moduli space of one $SO(10)$ instanton given by \eref{ExplicitJosephForSO2N-Symm} and \eref{ExplicitJosephForSO2N-Antisymm}, it is clear that here the rank-1 condition \eref{ExplicitJosephForSO2N-Antisymm} is missing altogether. As we have explained in the previous case, this can be at once read off from the relations \eref{FirstExplicitE6relns}-\eref{LastExplicitE6relns}, by keeping only the ones which are not corrected by instanton bilinears.
The classical dressing is a space of dimension 21 and again, by adding the contribution from the sum over instantons, we recover the correct 22-dimensional moduli space of one $E_6$ instanton.

\subsection{$N_f=6$}

The Hilbert series of $\widetilde{\CM}_{1, E_7}$ can be written as
\bea
H[\widetilde{\CM}_{1, E_7}](t; \vec x) = \sum_{n=0}^\infty {\Eseven n  0 0 0 0 0 0}_{\vec x} t^{2n}~.
\eea
The $E_7$ representations can be decomposed into those of $SO(12)\times U(1)$ using the projection matrix:
\bea
P_{E_7 \rightarrow D_6 \times U(1)} =\left(
\begin{array}{ccccccc}
 0 & 0 & 0 & 0 & 0 & 0 & 1 \\
 0 & 0 & 0 & 0 & 0 & 1 & 0 \\
 0 & 0 & 0 & 0 & 1 & 0 & 0 \\
 0 & 0 & 0 & 1 & 0 & 0 & 0 \\
 0 & 0 & 1 & 0 & 0 & 0 & 0 \\
 0 & 1 & 0 & 0 & 0 & 0 & 0 \\
 -2 & -2 & -3 & -4 & -3 & -2 & -1 \\
\end{array}
\right)~.
\eea
Under the action of this matrix, the fugacities $\vec x$ of $E_7$  are mapped to the fugacities $\vec y$ of $SO(12)$ and the fugacity $q$ of $U(1)$ as
\bea
\vec x = \left( \frac{1}{q^2},\frac{y_6}{q^2},\frac{y_5}{q^3},\frac{y_4}{q^4},\frac{y_3}{q^3},
\frac{y_2}{q^2},\frac{y_1}{q} \right)~.
\eea
We then have the following highest weight generating function:
\be
\begin{split}
&\cG[\widetilde{\CM}_{1, E_7}] (t; \mu_2, \mu_4, \mu_5 ; q) \\
&= \PE \Big[ \Big(1+ \mu_2 + \mu_5 (q + q^{-1})  + (q^2 + q^{-2}) \Big) t^2 + \mu_4 t^4 \Big]~,\label{E7dec1}
\end{split}
\ee
where at order $t^2$ we recognise the contributions of: $S$, which is a singlet of $SO(12)$; the instanton and the anti-instanton operators with $U(1)_I$ charge $\pm 1$ in the spinor representation $[0,0,0,0,1,0]$; the instanton and the anti-instanton operators with $U(1)_I$ charge $\pm 2$ which are singlets of $SO(12)$; the meson in the adjoint representation $[0,1,0,0,0,0]$. In addition there is a fourth-rank antisymmetric tensor of $SO(12)$ at order $t^4$.

\subsubsection{The generators and their relations}
The expansion up to order $t^4$ of \eref{E7dec1} is given by
\bea
& H[E_7](t; \vec x, q)  \nn \\
&= 1 +  \Big(1 + [0, 1, 0, 0, 0, 0] + (q+q^{-1}) [0, 0, 0, 0, 1, 0] + (q^2+q^{-2} ) \Big)t^2  \nn\\
&\qquad  +\Big(2+ [0, 2, 0, 0, 0, 0] +[0, 0, 0, 0, 2, 0] + [0, 0, 0, 1, 0, 0] + [0, 1, 0, 0, 0, 0] \nn \\
 & \qquad \quad   + (q+q^{-1}) (2 [0, 0, 0, 0, 1, 0] + [0, 1, 0, 0, 1, 0])  \nn \\
 & \qquad \quad +(q^{2}+q^{-2}) (1+[0, 0, 0, 0, 2, 0] + [0, 1, 0, 0, 0, 0]) \nn \\
 &  \qquad \quad+(q^{3}+q^{-3})[0, 0, 0, 0, 1, 0]  + (q^4+q^{-4}) \Big) t^4 +\ldots~. \label{HSE7Nf6}
\eea
The plethystic logarithm of this Hilbert series is given by
 \bea \label{plE7SO12}
&\PL \left[ H[E_7](t; \vec x, q) \right] \nn \\
&= \Big(1 + [0, 1, 0, 0, 0, 0] + (q+q^{-1}) [0, 0, 0, 0, 1, 0]  + (q^2+q^{-2} )\Big)t^2 -  \nn \\
& \quad - \Big( 2 + [0, 0, 0, 1, 0, 0] + [0, 1, 0, 0, 0, 0] +[2, 0, 0, 0, 0, 0] \nn \\
& \qquad + (q + q^{-1}) ([0, 0, 0, 0, 1, 0] + [1, 0, 0, 0, 0, 1]) + (q^2 + q^{-2}) [0, 1, 0, 0, 0, 0]   \Big) t^4 + \ldots~.
 \eea
  
For $SO(12)$, we use $a,b,c,d =1, \ldots, 12$ to denote vector indices, $\alpha, \beta, \rho, \sigma=1,\ldots, 32$ to denote indices of the spinor representation $[0,0,0,0,1,0]$, and $\dot \alpha, \dot \beta, \dot \rho, \dot \sigma=1,\ldots, 32$ to denote indices of the conjugate spinor representation $[0,0,0,0,0,1]$. The spinor representation of $SO(12)$ is pseudoreal, hence all contractions of the spinor indices are made with the epsilon tensor, which takes the forms
\bea
\epsilon_{\alpha \beta}  \quad \text{or} \quad \epsilon^{\alpha \beta}  \quad \text{or} \quad \epsilon_{\dot \alpha \dot \beta} \quad \text{or} \quad \epsilon^{\dot \alpha \dot \beta} ~.
\eea
Gamma matrices $\gamma^a$ take the forms
\bea
(\gamma^a)_{\alpha \dot \beta}~.
\eea
The product of two gamma matrices has the following forms:
\bea
(\gamma^{ab})_{\alpha \beta} \equiv (\gamma^{[a})_{\alpha \dot \alpha} (\gamma^{b]})_{ \beta \dot \beta} \epsilon^{\dot \alpha \dot \beta} \quad \text{and} \quad (\gamma^{ab})_{\dot \alpha \dot \beta} \equiv (\gamma^{[a})_{\alpha  \dot\alpha } (\gamma^{b]})_{ \beta \dot \beta} \epsilon^{\alpha \beta}~,
\eea
where the spinor indices are symmetric. 
The product of four gamma matrices has the following forms:
\bea
(\gamma^{abcd})_{\alpha\sigma} &\equiv (\gamma^{[a})_{\alpha\dot \alpha} (\gamma^b)_{\beta \dot \beta} (\gamma^c)_{\rho \dot \rho}  (\gamma^{d]})_{\sigma \dot \sigma} \epsilon^{\dot \alpha \dot \beta} \epsilon^{\beta \rho} \epsilon^{\dot \rho \dot \sigma} \\
(\gamma^{abcd})_{\dot \alpha \dot \sigma} &\equiv (\gamma^{[a})_{\alpha\dot \alpha} (\gamma^b)_{\beta \dot \beta} (\gamma^c)_{\rho \dot \rho}  (\gamma^{d]})_{\sigma \dot \sigma} \epsilon^{\alpha \beta} \epsilon^{\dot \beta \dot \rho} \epsilon^{ \rho \sigma} ~,
\eea
where the spinor indices are antisymmetric. 

The generators of the moduli space are $M^{ab}$, which is a $12 \times12$ antisymmetric matrix, the instanton operators $I_{1+}^{\alpha},I_{1-}^{\alpha}$ and $I_{2+}, I_{2-}$, and the glueball superfield $S$.

From \eref{plE7SO12}, we have the following sets of relations:
{\small
\bea
 [2,0,0,0,0,0]+[0,0,0,0,0,0]:&\qquad  M^{ab} M^{bc} =  (I^\alpha_{1+} \epsilon_{\alpha \beta} I^\beta_{1-} ) \delta^{ac} \label{rel1E7} \\ 
[0, 0, 0, 1, 0, 0]: & \qquad  M^{[a_1 a_2} M^{a_3 a_4]} = I^\alpha_{1+} {I}^\beta_{1-} (\gamma^{a_1 \cdots a_{4}})_{\alpha\beta} \\ 
[0,0,0,0,0,0]: & \qquad   S^2+ I_{2+} I_{2-}=I^\alpha_{1+} I^\beta_{1-} \epsilon_{\alpha \beta} \label{rel2E7}  \\
[0, 1, 0, 0, 0, 0]:& \qquad SM^{ab} = I^\alpha_{1+} {I}^\beta_{1-} (\gamma^{ab})_{\alpha\beta}  \\
(q^{2}+q^{-2})[0,1,0,0,0,0]: &\qquad     I_{2\pm}M^{ab} = I^\alpha_{1\pm} {I}^\beta_{1\pm} (\gamma^{ab})_{\alpha \beta} \\
(q +q^{-1})([1, 0, 0, 0, 0, 1]+[0, 0, 0, 0, 1, 0]): &\quad M^{ab} I^\alpha_{1\pm} (\gamma^b)_{\alpha \dot \beta} = (S I_{1\pm}^{\alpha}  + I_{2\pm }I_{1\mp}^{\alpha})  (\gamma^a)_{\alpha \dot \beta} ~.
\eea}
To aid computations it is useful to rewrite \eref{HSE7Nf6} and \eref{plE7SO12} in terms of characters of $SO(12) \times SU(2)$. The reader can find the relevant formulae in Appendix \ref{appSec:SO12xSU2}.

\subsubsection{Expansion in the instanton fugacity }

The highest weight generating function \eref{E7dec1} can be expanded in powers of the instanton number fugacity $q$ as
\be
\begin{split}
&\cG[\widetilde{\CM}_{1, E_7}](t; \mu_2, \mu_4, \mu_5 ; q) \\
&=\frac{1}{(1 - t^2)(1 - \mu_2 t^2)(1 - \mu_4 t^4)(1 - \mu_5^2 t^4)(1 - t^4)} \sum_{m \in \BZ} (t^2 \mu_5)^{|m|} q^m \sum_{n \in \BZ} t^{2 |n|} q^{2n}  \\      
&=\quad \PE[( \mu_5^2 +1+ \mu_2 )t^2+ \mu_4 t^4 + \mu_5^2 t^6] \\
&\quad \times \Big( \frac{1 + \mu_5^2 t^4}{ 1 - t^4} \sum_{m~\mathrm{even} } t^{|m|} q^m - \frac{(t \mu_5)^2}{  1 - \mu_5^2 t^4 } \sum_{m~\mathrm{even} } \mu_5^{|m|}t^{2|m|} q^m \\
&\quad + \frac{(1 +  t^2)\mu_5 t}{  1 - t^4 } \sum_{m~\mathrm{odd} } t^{|m|} q^m - \frac{(t \mu_5)^2}{  1 - \mu_5^2 t^4} \sum_{m~\mathrm{odd} } \mu_5^{|m|}t^{2|m|} q^m\Big)~.
\end{split}
\ee
The first equality is a $q$ expansion in terms of a double sum. This separates the classical dressing from the one and two instanton contributions. It is precisely the presence of both types of instantons as quadratic generators that, for $N_f>5$, complicates the features of the $q$ expansion in terms of a one sum only. We still write such an expansion in the second equality, splitting it into odd and even terms.

\subsection{$N_f=7$}

The Hilbert series of $\widetilde{\CM}_{1, E_8}$ can be written as
\bea \label{E8inst}
H[\widetilde{\CM}_{1, E_8}](t; \vec x) = \sum_{n=0}^\infty \left[ {\Eeight 0  0 0 0 0 0 0 n} \right]_{\vec x} t^{2n}~.
\eea
The $E_8$ representations can be decomposed into those of $SO(14)\times U(1)$ using the projection matrix
\bea
P_{E_8 \rightarrow D_7 \times U(1)} =\left(
 \begin{array}{cccccccc}
 0 & 0 & 0 & 0 & 0 & 0 & 0 & 1 \\
 0 & 0 & 0 & 0 & 0 & 0 & 1 & 0 \\
 0 & 0 & 0 & 0 & 0 & 1 & 0 & 0 \\
 0 & 0 & 0 & 0 & 1 & 0 & 0 & 0 \\
 0 & 0 & 0 & 1 & 0 & 0 & 0 & 0 \\
 0 & 0 & 1 & 0 & 0 & 0 & 0 & 0 \\
 0 & 1 & 0 & 0 & 0 & 0 & 0 & 0 \\
 -4 & -5 & -7 & -10 & -8 & -6 & -4 & -2 \\
\end{array}
\right)~.
\eea
Under the action of this matrix, the fugacities $\vec x$ of $E_8$  are mapped to the fugacities $\vec y$ of $SO(14)$ and the fugacity $q$ of $U(1)$ as
\bea
\vec x = \left( \frac{1}{q^4},\frac{y_7}{q^5},\frac{y_6}{q^7},\frac{y_5}{q^{10}},\frac{y_4}{q^8},\frac{y_3}{q^6},\frac{y_2}{q^4},\frac{y_1}{q^2} \right)~.
\eea
We then have the following highest weight generating function:
\be\label{HWGE8}
\begin{split}
\cG[\widetilde{\CM}_{1, E_8}] (t; \vec \mu ; q)&= \PE \Bigg[t^2 \left(1+\mu _2+\mu _6 q+\mu _7q^{-1} +\mu _1 (q^2+q^{-2}) \right) \\
& \qquad + t^4 \left(1+\mu _2+\mu _4 +\mu _6 q+\mu _7q^{-1} +\mu _3 (q^2+q^{-2})\right)   \\
& \qquad + t^6 \left(\mu _4+\mu _5 (q^2+q^{-2})\right)\Bigg]~.
\end{split}
\ee

\subsubsection{The generators and their relations}

The Hilbert series of the reduced moduli space of $1$ $E_8$ instanton can be written in terms of characters of $SO(14) \times U(1)$ as
\bea
& H[E_8](t; \vec x, q)  \nn \\
&= 1 +  \Big(  (1 + [0, 1, 0, 0, 0, 0, 0]) + [0, 0, 0, 0, 0, 1, 0] q + [0, 0, 0, 0, 0, 0, 1]q^{-1}  \nn \\
& \qquad  + [1, 0, 0, 0, 0, 0, 0] (q^2+q^{-2}) \Big)t^2+ \Big( 2 + 
   [0, 0, 0, 0, 0, 1, 1] + [0, 0, 0, 1, 0, 0, 0] \nn \\
   &\qquad + 2 [0, 1, 0, 0, 0, 0, 0] + [0, 2, 0, 0, 0, 0, 0]  + [2, 0, 0, 0, 0, 0, 0]  \nn \\
 &\qquad  + (2 [0, 0, 0, 0, 0, 1, 0] + 
    [0, 1, 0, 0, 0, 1, 0] + 
    [1, 0, 0, 0, 0, 0, 1]) q \nn \\
  &\qquad + (2 [0, 0, 0, 0, 0, 0, 1] + 
   [0, 1, 0, 0, 0, 0, 1] + [1, 0, 0, 0, 0, 1, 0]) q^{-1} \nn \\  
  &\qquad + ([0, 0, 0, 0, 0, 2, 0] + 
    [0, 0, 1, 0, 0, 0, 0] + [1, 0, 0, 0, 0, 0, 0] + 
    [1, 1, 0, 0, 0, 0, 0]) q^2 \nn \\
    & \qquad+ ([0, 0, 0, 0, 0, 0, 2] + [0, 0, 1, 0, 0, 0, 0] +
    [1, 0, 0, 0, 0, 0, 0] + 
   [1, 1, 0, 0, 0, 0, 0])q^{-2} \nn \\
  &\qquad + [1, 0, 0, 0, 0, 1, 0] (q^3+q^{-3}) + 
 [2, 0, 0, 0, 0, 0, 0] (q^4+q^{-4}) \Big) t^4 +\ldots~. \label{HSE8Nf7}
\eea
The plethystic logarithm of this Hilbert series is given by
 \bea \label{plE8SO14}
&\PL \left[ H[E_8](t; \vec x, q) \right] \nn \\
&=  \Big(  (1 + [0, 1, 0, 0, 0, 0, 0]) + [0, 0, 0, 0, 0, 1, 0] q + [0, 0, 0, 0, 0, 0, 1]q^{-1}  \nn \\
& \qquad  + [1, 0, 0, 0, 0, 0, 0] (q^2+q^{-2}) \Big)t^2 - \Big( 2+ [2, 0, 0, 0, 0, 0, 0] + [0, 0, 0, 1, 0, 0, 0] + [0, 1, 0, 0, 0, 0, 0]  \nn \\
 & \qquad +  ([0, 0, 0, 0, 0, 1, 0] + [1, 0, 0, 0, 0, 0, 1]) q +  ([0, 0, 0, 0, 0, 0,1] + [1, 0, 0, 0, 0, 1, 0]) q^{-1} \nn \\ 
 & \qquad + ([0, 0, 1, 0, 0, 0, 0] + [1, 0, 0, 0, 0, 0, 0]) (q^2+q^{-2})  \nn \\
 & \qquad + [0, 0, 0, 0, 0, 0, 1]q^3  +  [0, 0, 0, 0, 0, 1, 0] q^{-3} +(q^4+q^{-4})\Big)t^4 + \ldots~.
\eea

It is also useful to write the Hilbert series written in terms of characters of representations of $SO(16)$:
\bea
& H[E_8](t; \vec z)  \nn \\
&= 1 +  ([0, 0, 0, 0, 0, 0, 0, 1] + [0, 1, 0, 0, 0, 0, 0, 0])t^2 +  \nn \\
& \qquad  (1 + [0, 0, 0, 0, 0, 0, 0, 1] + 
    [0, 0, 0, 0, 0, 0, 0, 2]  \nn \\
& \qquad + [0, 0, 0, 1, 0, 0, 0, 0] + 
    [0, 1, 0, 0, 0, 0, 0, 1] + [0, 2, 0, 0, 0, 0, 0, 0])t^4 \ldots~. \label{HSSO16}
\eea
The plethystic logarithm of this Hilbert series is
\be
\begin{split}
&\PL[H[E_8](t; \vec z)] \\
& = ([0, 0, 0, 0, 0, 0, 0, 1] + [0, 1, 0, 0, 0, 0, 0, 0])t^2 - \Big( 1 + [0, 0, 0, 1, 0, 0, 0, 0] \\
&\quad + [1, 0, 0, 0, 0, 0, 1, 0] + 
 [2, 0, 0, 0, 0, 0, 0, 0] \Big) t^4 +\ldots~.
\end{split}
\ee

Note that the spinor representation $[0, 0, 0, 0, 0, 0, 0, 1]$ of $SO(16)$ branches to those of $SO(14) \times U(1)$ as 
\bea
[0, 0, 0, 0, 0, 0, 0, 1] \longrightarrow [0, 0, 0, 0, 0, 0, 1]_{-1} + [0, 0, 0, 0, 0, 1, 0]_{+1}  ~,
\eea
corresponding to the charge $\pm 1$ instanton operators $I_{1-}$ and $I_{1+}$, whereas the field $X$ in the adjoint representation $[0, 1, 0, 0, 0, 0, 0, 0]$ of $SO(16)$ contains the charge $\pm 2$ instanton operators  $I_{2+}$, $I_{2-}$, the glueball superfields $S$ and the meson $M$.

Thus, one independent singlet at order $t^4$ of $\eref{HSSO16}$ implies that $I_{1+} I_{1-}$ is proportional to the singlet formed by $I_{2+}$, $I_{2-}$, $S$ and $M$ in $X$.
The adjoint field $X$ of $SO(16)$ satisfies the matrix relation 
\bea X^2=0~,\label{sqX0} \eea
transforming in the rank two symmetric representation $[2,0,0,0,0,0,0,0]+ [0,0,0,0,0,0,0,0] $ of $SO(16)$.  This representation branches into those of $SO(14)\times U(1)$ as 
\bea
 [2,0,0,0,0,0,0,0] &\longrightarrow 1+[0,0,0,0,0,0,0]_{-4} +[0,0,0,0,0,0,0]_{+4} \nn \\
 & \qquad +[1,0,0,0,0,0,0]_{-2} +[1,0,0,0,0,0,0]_{+2} +[2,0,0,0,0,0,0]_{0}.
\eea
Upon expanding \eref{sqX0} in components, we see that the vanishing components $(X^2)_{15,15}$, $(X^2)_{16,16}$ and $(X^2)_{15,16}$ imply that 
\bea
I^a_{2+} I^a_{2+} =0~,\qquad  I^a_{2-} I^a_{2-} =0~, \qquad S^2+I^a_{2+} I^a_{2-} =0~.
\eea
These relations are collected in \eref{SI2I2} and \eref{II0E8}.

For future reference, the branching rule of the representation $[1,0,0,0,0,0,1,0]$ of $SO(16)$ to those of $SO(14) \times U(1)$ is
\bea
[1,0,0,0,0,0,1,0] &\longrightarrow [0,0,0,0,0,0,1]_{-3} +[0,0,0,0,0,0,1]_{+1} +[0,0,0,0,0,1,0]_{-1} \nn \\
& \quad +[0,0,0,0,0,1,0]_{+3} +[1,0,0,0,0,0,1]_{-1} +[1,0,0,0,0,1,0]_{+1}~,
\eea
and the branching rule of the representation $[0,0,0,1,0,0,0,0]$ of $SO(16)$ is
\bea
[0,0,0,1,0,0,0,0] &\longrightarrow [0,0,0,1,0,0,0]_{0} +[0,0,1,0,0,0,0]_{-2} +[0,0,1,0,0,0,0]_{+2} \nn \\
&\quad +[0,1,0,0,0,0,0]_{0}~.
\eea   

For $SO(14)$, we use $a,b,c,d =1, \ldots, 14$ to denote vector indices and $\alpha, \beta, \rho, \sigma=1,\ldots, 64$ to denote the spinor indices. Note that the spinor representation of $SO(14)$ is complex.  The delta symbol has the form
\bea
\delta^\alpha_\beta~.
\eea
The gamma matrices $\gamma^a$ can take the following forms:
\bea
(\gamma^a)_{\alpha  \beta}  \qquad \text{or} \qquad (\gamma^a)^{\alpha \beta}~,
\eea
where the $\alpha, \beta$ indices are antisymmetric.  The product of two gamma matrices is
\bea
(\gamma^{ab})^{\alpha}_{~\rho} \equiv (\gamma^{[a})^{\alpha\beta} (\gamma^{b]})_{\beta \rho} ~.
\eea
The product of three gamma matrices has the forms
\bea
(\gamma^{abc})_{\alpha \rho} \equiv (\gamma^{[a})_{\alpha\beta} (\gamma^b)^{\beta \sigma} (\gamma^{c]})_{\sigma \rho} \qquad \text{and} \qquad (\gamma^{abc})^{\alpha \rho} \equiv (\gamma^{[a})^{\alpha\beta} (\gamma^b)_{\beta \sigma} (\gamma^{c]})^{\sigma \rho}~,
\eea
symmetric in the spinor indices.  The product of four gamma matrices is
\bea
(\gamma^{abcd})^{\alpha}_{~ \beta} &\equiv (\gamma^{[a})^{\alpha\sigma_1} (\gamma^b)_{\sigma_1\sigma_2} (\gamma^c)^{\sigma_2\sigma_3}  (\gamma^{d]})_{\sigma_3 \beta} ~.
\eea
The generators of the moduli space are $M^{ab}$, which is a $14 \times14$ antisymmetric matrix;  the instanton operators $I^{\alpha}$ and $\widetilde{I}_{\alpha}$; and the gaugino superfield $S$.

The relations corresponding to order $t^4$ of \eref{plE8SO14} are as follows:
{\small
\bea
 [2, 0, 0, 0, 0, 0, 0]+[0, 0, 0, 0, 0, 0, 0]: & ~ M^{ab} M^{bc} +I^{(a}_{2+} I^{c)}_{2-} =  I^\alpha_{1+} (I_{1-})_\alpha\delta^{ac}     \label{sym2D6}\\
 [0, 0, 0, 1, 0, 0, 0]: & ~ M^{[a_1 a_2} M^{a_3 a_4]} = (I_{1-})_\beta (\gamma^{a_1 \cdots a_{4}} )^\beta_{~\alpha} I^\alpha_{1+} \\
   [0, 0, 0, 0, 0, 0, 0]: & ~ S^2+ I^a_{2+} I^a_{2-} =0 \label{SI2I2}\\ 
   [0, 1, 0, 0, 0, 0, 0]: & ~ SM^{ab}+ I^{[a}_{2+} I^{b]}_{2-}=  I^\alpha_{1+} (I_{1-})_\beta (\gamma^{ab})^{\beta}_{~\alpha}\\   
 q ([0, 0, 0, 0, 0, 1, 0] + [1, 0, 0, 0, 0, 0, 1]): & ~   M^{ab} I^\alpha_{1+} (\gamma^b)_{\alpha\beta}  \nn \\
 & \qquad =S I^\alpha_{1+} (\gamma^a)_{\alpha \beta} + I^a_{2+}(I_{1-})_\beta\\  
  q^{-1} ([0, 0, 0, 0, 0, 0, 1]+[1, 0, 0, 0, 0, 1, 0]): & ~   M^{ab} (I_{1-})_\alpha (\gamma^b)^{\alpha\beta} \nn \\
  & \qquad =S (I_{1-})_\alpha (\gamma^a)^{\alpha \beta} + I^a_{2-}I^\beta_{1+}\\   
q^2 [0,0,1,0,0,0,0]: & ~  M^{[ab} I^{c]}_{2+} = I^\alpha_{1+}(\gamma^{a b c})_{\alpha \beta}  I^\beta_{1+}   \\
q^{-2} [0,0,1,0,0,0,0]: & ~  M^{[ab} I^{c]}_{2-} = (I_{1-})_\alpha (\gamma^{a b c})^{\alpha \beta}  (I_{1-})_\beta   \\
q^2 [1, 0, 0, 0, 0, 0, 0]: & ~  M^{ab} I^b_{2+}= S I^a_{2+} \\
q^{-2} [1, 0, 0, 0, 0, 0, 0]: & ~  M^{ab} I^b_{2-}= S I^a_{2-} \\
 q^3 [0, 0, 0, 0, 0, 0, 1]: & ~I^a_{2+} I^\alpha_{1+} (\gamma^a)_{\alpha \beta} =0 \\
  q^{-3} [0, 0, 0, 0, 0, 0, 1]: & ~I^a_{2-} (I_{1-})_\alpha (\gamma^a)^{\alpha \beta} =0 \\
  (q^4+q^{-4})[0, 0, 0, 0, 0, 0, 0]: &~  I^a_{2+} I^a_{2+} = I^a_{2-} I^a_{2-} = 0~. \label{II0E8}
\eea}

\subsubsection{Expansion in the instanton fugacity }

The highest weight generating function \eref{HWGE8} can be rewritten in terms of an implicit expansion in $q$ involving 5 sums: 
\be 
\begin{split}
&\cG[\widetilde{\CM}_{1, E_8}] (t; \vec \mu ; q)= \PE \Big[ \left(1+\mu _2\right) t^2 + \left(1+\mu _2 + \mu _4 \right)t^4+\mu_4 t^6\Big] \\
& ~~~ \times \PE \Big[ (\mu _6 \mu_7  + \mu_1^2) t^4+ (\mu _6 \mu_7  +\mu_3^2) t^8 +\mu_5^2 t^{12}  \Big] \\
& ~~~ \times \Big(\sum_{n_1 \geq 0} ( \mu_6 t^2 q)^{n_1}   + \sum_{n_1<0} ( \mu_7 t^2 )^{-n_1} q^{n_1} \Big)  \sum_{n_2 \in \BZ } ( \mu_1 t^2)^{|n_2|} q^{2n_2}  \\
& ~~~ \times \Big(\sum_{n_3 \geq 0} ( \mu_6 t^4 q)^{n_3}   + \sum_{n_3<0} ( \mu_7 t^4 )^{-n_3} q^{n_3} \Big)  \sum_{n_4 \in \BZ } ( \mu_3 t^4)^{|n_4|} q^{2 n_4} \sum_{n_5 \in \BZ } ( \mu_5 t^6)^{|n_5|} q^{2 n_5} ~.
\end{split}
\ee

\section{$USp(4)$ with one antisymmetric hypermultiplet}\label{sec:USp4}

In this theory, we pick the trivial value of the discrete theta angle for the $USp(4)$ gauge group.  The Higgs branch at infinite coupling of this theory is identified with the reduced moduli space of 2 $SU(2)$ instantons on $\BC^2$ \cite{Intriligator:1997pq}, whose global symmetry is $SU(2) \times SU(2)$.  The Hilbert series is given by (3.14) of \cite{Hanany:2012dm}.  For reference, we provide here the explicit expression of the Hilbert series up to order $t^6$:
\bea
H[\widetilde{\CM}_{2, SU(2)}] (t; y,x )&=1 + ([0; 2] + [2; 0]) t^2 + 
 [1; 2] t^3 + (1 + [0; 4] + [2; 2] + [4; 0]) t^4 \nn \\
 & \qquad + ([1; 2] + 
    [1; 4] + [3; 2]) t^5 + ([0; 2] + [0; 6] + [2; 0]  \nn \\
 &\qquad    +2 [2; 4] + [4; 2] + [6; 0]) t^6 +\ldots~.
\eea
The plethystic logarithm of this expression is
\be
\begin{split}
\PL \left[ H[\widetilde{\CM}_{2, SU(2)}] (t; y,x ) \right] &= ([0; 2] + [2; 0]) t^2 + 
 [1; 2] t^3 - t^4 -([1;2]+[1;0])t^5  \\
 & \quad - ([2;0]+[0;2])t^6+ \ldots~.
\end{split}
\ee
The corresponding highest weight generating function is (see (4.25) of \cite{Hanany:2014dia})
\bea
\cG[\widetilde{\CM}_{2, SU(2)}] (t; \mu_1, \mu_2)= \PE \left[ (\mu_1^2 + \mu_2^2) t^2 + \mu_1 \mu_2^2 t^3 + t^4 + \mu_1 \mu_2^2 t^5 - \mu_1^2 \mu_2^4 t^{10} \right]~,
\eea
where $\mu_1$ and $\mu_2$ are respectively the fugacities for the highest weights of the $SU(2)$ acting on the centre of instantons and the $SU(2)$ associated with the internal degrees of freedom.

Let us use the indices $a,b,c,d=1,2$ for the first $SU(2)$ and $i,j,k,l =1,2$ for the second $SU(2)$. The generators of the moduli space are as follows.  
\bi
\item {\bf Order $t^2$:} The rank two symmetric tensors $P_{ab}$ and $M_{ij}$ in the representation $[2;0]$ and $[0;2]$ of $SU(2) \times SU(2)$:
\bea
P_{ab} = P_{ba}~, \qquad M_{ij} = M_{ji}~.
\eea
\item {\bf Order $t^3$:} A doublet of rank two symmetric tensors $(A_a)_{ij}$, with
\bea
(A_a)_{ij} = (A_a)_{ji}~,
\eea
in the representation $[1;2]$ of $SU(2) \times SU(2)$.
\ei
The singlet relation at order $t^4$ can be written as
\bea
[0;0]t^4: \qquad \Tr(P^2) = \Tr(M^2)~.
\eea
The relations at order $t^5$ are
\bea
[1;0]t^5: &\qquad \epsilon^{ii'} \epsilon^{jj'} (A_a)_{ij} M_{i'j'} = 0~, \\
[1;2]t^5:  &\qquad \epsilon^{bb'} P_{ab} (A_{b'})_{ij}= \epsilon^{kk'} M_{ik}( A_{a})_{k'j} + (i \leftrightarrow j)~.
\eea
The relations at order $t^6$ are
\bea
[2;0]t^6: &\qquad \Tr(P^2) P_{ab}  =\epsilon^{ii'} \epsilon^{jj'}(A_a)_{ij} (A_b)_{i'j'}~, \\
[0;2]t^6:  &\qquad \Tr(M^2) M_{ij} = \epsilon^{ab}\epsilon^{kk'} (A_a)_{ik} (A_b)_{k'j} ~.
\eea

Let us now rewrite the above statements in $SU(2) \times U(1)$ language. Up to charge normalisation, we identify the Cartan subalgebra of the latter $SU(2)$ associated with $\mu_2$ with the $U(1)_I$ symmetry. More precisely, if $w$ is the fugacity associated to the Cartan generator of the latter $SU(2)$, then $q=w^2$ is the fugacity for the topological symmetry. The highest weight generating function can then be written as 
\be
\begin{split}
\cG[\widetilde{\CM}_{2, SU(2)}] (t; \mu_1; q) &= \PE \Big[ \Big(1 + \mu_1^2 + (q+q^{-1}) \Big) t^2 + \Big(\mu_1 + \mu_1 (q+q^{-1}) \Big) t^3  \\
& \qquad - \mu_1 t^5 - \mu_1^2 t^6 \Big]~.
\end{split}
\ee
This can be written as a power series in $q$ as 
\be
\begin{split}
\cG[\widetilde{\CM}_{2, SU(2)}] (t; \mu_1; q) =\frac{1}{\left(1-t^2\right) \left(1-t^4\right) (1-\mu_1 t) \left(1-\mu_1^2 t^2\right) \left(1-\mu_1 t^3\right)} \times \\
 \Big( (1-\mu_1^2 t^6) \sum_{j=-\infty}^\infty q^j t^{2 |j|} - \left(1-t^4\right)\sum_{j=-\infty}^\infty q^j t^{2 |j|} (\mu_1 t)^{|j|+1}  \Big)~.
\end{split}
\ee
The Hilbert series up to order $t^6$ can be written explicitly as follows:
\be
\begin{split}
H[\widetilde{\CM}_{2, SU(2)}](t; y,q) &= 1 + \Big(1+[2]+(q+q^{-1}) \Big) t^2 + \Big( [1]+ [1](q+q^{-1}) \Big) t^3  \\
& + \Big(2+[2]+[4]+ (1+[2])(q+q^{-1})+(q^2+q^{-2}) \Big) t^4  \\
& + \Big(2 [1]+[3]+(2  [1]+ [3])(q+q^{-1})+ [1](q^2+q^{-2}) \Big) t^5  \\
& + \Big(2+3 [2]+[4]+[6]+(2+2  [2]+ [4])(q+q^{-1}) \\
& \qquad +(1+2  [2])(q^2+q^{-2})+(q^3+q^{-3}) \Big) t^6 + \ldots~.
\end{split}
\ee
The plethystic logarithm of this Hilbert series is given by
\bea
\PL \left[ H[\widetilde{\CM}_{2, SU(2)}](t; y,q) \right]&= \Big(1+[2]+(q+q^{-1}) \Big) t^2 + \Big( [1]+ [1](q+q^{-1}) \Big) t^3  -t^4  \nn \\
&  \quad -\Big(2[1]+[1](q+q^{-1}) \Big) t^5- \Big(1+[2]+(q+q^{-1}) \Big) t^6 \nn \\
& \quad + \ldots~.
\eea

\paragraph{The generators.} At order $t^2$, the generators are 
\bea
[2]: &\qquad \qquad P_{ab}  \quad  \text{with $P_{ab} = P_{ba}$}~,\\
q,\; q^{-1},\; 1: &\qquad  \qquad I, \; \widetilde{I}, \; S~.
\eea
The generators $P_{ab}$ are identified as a product of two antisymmetric tensors:
\bea
P_{ab}  = \Tr(X_a X_b)~.
\eea
At order $3$, the generators are denoted by
\bea
q[1], \; q^{-1}[1],\; [1]:  \qquad J_{a}~, \quad \widetilde{J}_{a}~, \quad T_a~.
\eea
where the generators $T_a$ are identified as a product of two gauginos and one antisymmetric tensor
\bea
T_a = \Tr \left( X_{a} \mathcal{W} \mathcal{W} \right)~.
\eea

\paragraph{The relations.} The relation at order $t^4$ can be written as
\bea
[0]t^4: \qquad  \Tr(P^2)+S^2 = I \widetilde{I}~.
\eea
The relations at order $t^5$ can be written as
\bea
[1]t^5: &\qquad    ST_a =\widetilde{I}  J_{a} + I \widetilde{J}_a~,\\
q[1] t^5: &\qquad  P_{ab} J_{b'} \epsilon^{bb'} + I T_a + S {J_a} =0~,\\
[1]t^5: &\qquad  P_{ab} T_{b'}\epsilon^{bb'} +2S T_a =0~,  \\
q^{-1}[1]t^5: & \qquad P_{ab} \widetilde{J}_{b'}\epsilon^{bb'} + \widetilde{I} T_a + S \widetilde{J_a}=0~.
\eea
The relations at order $t^6$ can be written as
\bea
[2]t^6: &\qquad S^2P_{ab}  +T_{a}T_{b} =  J_{(a} \widetilde{J}_{b)} + I \widetilde{I} P_{ab}~, \\
q t^6: &\qquad S^2 I =  \epsilon^{a b} J_a T_{b}+I^2 \widetilde{I}~, \\
t^6: &\qquad S^3 = \epsilon^{a b} J_a \widetilde{J}_{b}+S I \tilde{I}~, \\
q^{-1} t^6: &\qquad S^2 \widetilde{I} =  \epsilon^{a b} \widetilde{J}_a T_{b}+ \widetilde{I}^2 I~.
\eea

\section{$USp(2k)$ with one antisymmetric hypermultiplet}\label{sec:USp2k}

As in the previous sections, we pick the trivial value of the discrete theta angle for $USp(2k)$ gauge group.  The Higgs branch of the conformal field theory at infinite coupling is identified with the moduli space of $k$ $SU(2)$ instantons on $\BC^2$ \cite{Intriligator:1997pq}.  Below we consider the moduli space of the theory at finite coupling.

For $k=1$, the Higgs branch at finite coupling is
\bea
\BC^2 \times \mathbb{Z}_2~,
\eea
where $\BC^2$ is the classical moduli space of a $USp(2)$ gauge theory with 1 antisymmetric hypermultiplet and $\mathbb{Z}_2$ is the moduli space generated by the glueball superfield $S$ such that $S^2=0$. The Hilbert series is then given by
\be
\begin{split}
H_{k=1}(t;x,w) &= H[\mathbb{Z}_2] (t; w) H[\BC^2](t;x) \\
&= (1+w^2 t^2) \PE \left[ t(x+x^{-1}) \right] = \frac{1+w^2 t^2}{(1-t x)(1-t x^{-1})}~,
\end{split}
\ee
where the fugacity $w$ corresponds to the number of gaugino superfields. 

For higher $k$, the theory in question can be realised as the worldvolume theory of $k$ coincident D4-branes on an $O8^-$ plane.  Hence, the moduli space is expected to be the $k$-th symmetric power of $\BC^2 \times \bZ_2$, whose Hilbert series is given by
\be
\begin{split}
H_k (t,x,w) &= \oint_{|\nu|=1} \frac{{\rm d} \nu}{2 \pi i \nu^{k+1}} \exp \left(\sum_{m=1}^\infty \frac{\nu^m}{m} H_{k=1}(t^m;x^m,w^m) \right) \\
&= \sum_{j=0}^k (wt)^{2j} H[\Sym^j \BC^2](t, x) H[\Sym^{k-j} \BC^2](t,x)~,
\end{split}
\ee
where $H[\Sym^n \BC^2](t,x)$ is the Hilbert series for the $n$-th symmetric power of $\BC^2$:
\bea
H[\Sym^n \BC^2](t,x)= \oint_{|\nu|=1} \frac{{\rm d} \nu}{2 \pi i \nu^{n+1}} \exp \left(\sum_{m=1}^\infty \frac{\nu^m}{m} \frac{1}{(1-t^m x^m)(1-t^m x^{-m})} \right)~.
\eea
 We tested the result for $k=2$ directly from the field theory side using {\tt Macaulay2}; the details are presented in Appendix \ref{appSec:chiralRingHSwithGaugino}.
 
Note that this result also holds for $USp(2k)$ gauge theory with 1 antisymmetric hypermultiplet and 1 fundamental hypermultiplet.  This is because the classical moduli space of this theory is the moduli space of $k$ $SO(2)$ instantons on $\BC^2$ --- this space is in fact the $k$-symmetric power of the moduli space of $1$ $SO(2)$ instanton on $\BC^2$, which is identical to $\BC^2$.

Since the symmetric product $\Sym^k (\BC^2 \times \mathbb{Z}_2)$ has a $\BC^2$ component that can be factored out, it is natural to define the Hilbert series $\widetilde{H}_k(t;x,w)$ of the reduced moduli space  as follows:
\bea
H_k(t;x,w) = H[\BC^2](t;x)  \widetilde{H}_k(t;x,w)  = \frac{1}{(1-tx)(1-t x^{-1})}  \widetilde{H}_k(t;x,w)~.
\eea

\paragraph{Examples.} For $k=2$, we have
\bea \label{USp41antisym}
\widetilde{H}_{k=2} (t,x,w) &= (1 + w^4 t^4) (1 - 
     t^4) \PE[(x^2 + 1 + x^{-2})  t^2] + (w t)^2  \PE[(x + x^{-1}) t]\nn \\
&= 1+([2]+w^2) t^2 + [1] w^2 t^3 +([4] +  [2] w^2 + w^4)t^4+([3]w^2)t^5 \nn \\
& \qquad +([6]+[4]w^2+[2]w^4)t^6 +\ldots~.
\eea
The plethystic logarithm of this Hilbert series is
\be
\begin{split}
&\PL[\widetilde{H}_{k=2} (t,x,w)] \\
&= ([2]+w^2)t^2 +[1]w^2 t^3 -t^4-[1](w^2+w^4)t^5- ([2]w^4+w^6)t^6 +\ldots~.
\end{split}
\ee
For $k=3$, we have
\be
\begin{split}
\widetilde{H}_{k=3} (t,x,w) &= 1+([2]+w^2) t^2 + ([3]+[1] w^2) t^3 +  (1+[4] + 2 [2]w^2  + 
    w^4 ) t^4 \\
    & \quad + ([3]+[5] +([1]+2[3]) w^2   +  [1]w^4) t^5 + \Big([2]+2 [6] \\
 & \quad + (1+[2]+3  [4])w^2 + 2  [2]w^4  + w^6 \Big) t^6 +\ldots~.
\end{split}
\ee
The plethystic logarithm of this Hilbert series is
\be
\begin{split}
\PL[\widetilde{H}_{k=3} (t,x,w)] &= ([2]+w^2) t^2 + ([3]+[1] w^2) t^3 +[2]w^2  t^4 - [1] t^5 \\
& \qquad - \Big( [2]+(1+  [2] )w^2+ [2] w^4 \Big) t^6 +\ldots~.
\end{split}
\ee

\paragraph{General $k$.} For general $k$, we have two sets of generators transforming in:
\ben
\item representation $[p]$ at order $t^p$, for all $2\leq p \leq k$;
\item representation $[p] w^2$ at order $t^{p+2}$, for all $0\leq p \leq k-1$;
\een
these follow from the generators of the moduli space of two instantons, given by section 8.5 of \cite{Cremonesi:2014xha}.  Explicitly, these generators are
\bea \label{USp2kgen}
&\Tr(X_{a_1}X_{a_2}), \Tr(X_{a_1} X_{a_2} X_{a_3}), \ldots~, \Tr(X_{a_1} X_{a_2} \cdots X_{a_k}),\\
&\Tr({\cal W} {\cal W}), \;  \Tr(X_{a_1} {\cal W} {\cal W}), \; \Tr(X_{(a_1} X_{a_2)} {\cal W} {\cal W})~, \ldots, \; \Tr(X_{(a_1} \cdots X_{a_{k-1})} {\cal W} {\cal W}) \nn
\eea
where $a_1, a_2, \ldots, a_k =1,2$.  The set of relations with the lowest dimension transform in the representation $[k-2]$ at order $t^{k+2}$. 

In the limit $k \rightarrow \infty$, the moduli space is thus freely generated by \eref{USp2kgen}.%
\footnote{We would like to express our thanks to Nick Dorey for his nice presentation at the Swansea workshop and especially for discussing this point.}  A similar situation was considered in \cite{Benvenuti:2006qr}, where it was pointed out that the generating function of multi-trace operators for one brane is equal to that of single trace operators for infinitely many branes.

\section{Pure super Yang-Mills theories}\label{sec:SUN}
For 5d $\CN=1$ pure Yang-Mills theory, the Higgs branch at infinite coupling takes a simple orbifold structure. Field theoretic and stringy arguments can be provided for this statement.

In \cite{Tachikawa:2015mha} it was argued by counting zero modes that for an $SU(N)$ gauge group the instanton operators transform in the spin-$\frac{N}{2}$ representation of $SU(2)_R$. In \cite{Zafrir:2015uaa} the result was generalised to arbitrary gauge groups. Using the observation of \cite{Rodriguez-Gomez:2013dpa} that the instanton contribution to the superconformal index is given by an ``$SU(2)$-covariantized" version of the Hilbert series, the $SU(2)_R$ spin of instanton operators in pure Yang-Mills theories is given by $\frac{1}{2}h^\vee_G$, where $h^\vee_G$ is the dual Coxeter number of the group $G$. 
It is then straightforward to construct the relation between the instanton operators and the glueball operator:
\bea
S^{h^\vee_G}=I \tI ~.
\eea
which reduces to the standard nilpotency for $S$ \cite{Cachazo:2002ry} at finite coupling where the instanton operators are set to zero. The Higgs branch at infinite coupling is thus the orbifold $\BC^2/\BZ_{h^\vee_G}$.

For $SU(N)$ pure Yang-Mills a stringy construction provides a complementary viewpoint. For this therory, an $SL(2,\BZ)$ transformation on the 5-brane web can be exploited to set the charges of the external 5-brane legs to $(p_1, q_1)= (N,-1)$ and $(p_2, q_2) =(0,1)$.  In this basis, the web can be depicted as follows (this example is for $N=3$): 
\bea
\begin{tikzpicture}[scale=0.8, transform shape]
\draw (0,-1)--(0,0); 
\draw (0,0)--(4,0); 
\draw (0,0)--(-1,1); 
\draw (-1,1)--(2,1); 
\draw (4,0)--(7,-1); 
\draw (-1,1)--(-3,2);
\draw (2,1)--(4,0); 
\draw (1,2)--(2,1); 
\draw (-3,2)--(1,2); 
\draw (-3,2)--(-6,3); 
\draw (1,2)--(1,3); 
\end{tikzpicture}
\eea
At infinite coupling, the two 5-branes intersect and move apart, giving a one quaternionic dimensional Higgs branch, which has a cone structure. Using the classification of hyperK\"ahler cones of dimension 1, the space has to be an ADE singularity. The existence in the chiral ring of the operator $S$, which has spin-1 under $SU(2)_R$, rules out the D and E cases, implying that the Higgs branch has to be $\BC^2/\BZ_m$, for some $m$. The value of $m$ can be deduced by considering the intersection number, which is given by:
\bea
p_1 q_2 - p_2 q_1 = N~.
\eea
The Higgs branch at infinite coupling is therefore $\BC^2/\BZ_N$.%
\footnote{We thank Cumrun Vafa for discussions about this point.} 
%Separating the intersection points leads to resolution of the singularity.

The generators of the Higgs branch at infinite coupling are $I$, $S$, $\widetilde{I}$, singlets under $SU(N)$, and with $U(1)_I$ charge $+1,~0~\mathrm{and}~-1$ respectively. For $N>2$, the isometry group of $\BC^2/\BZ_N$ is identified with $U(1)_I$. For $N=2$, the isometry of the Higgs branch is enhanced to $SU(2)$ and the operators form a triplet $(I,S,\widetilde{I})$.

The construction can be generalised by means of orientifold planes \cite{Aharony:1997bh} to give analogous results for the case of classical gauge groups.

\section{Discussion}\label{sec:Discussion}

A coherent picture of the Higgs branch of 5d $\mathcal{N}=1$ theories for all values of the gauge coupling emerges from this paper. In particular, we have presented explicit relations that define the chiral ring at infinite coupling and are consistent with those at finite coupling. A crucial result of this paper is the correction to the glueball superfield, $S$, which at finite coupling is a nilpotent bilinear in the gaugino superfield and at infinite coupling becomes an ordinary chiral operator on the Higgs branch. 

For pure $SU(2)$ theories with $N_f \leq 7$ flavours a nice pattern was established. The finite coupling relations involving mesons and the glueball operator are corrected at infinite coupling by bilinears in the instanton operators, in the obvious way dictated by representation theory. New relations also arise which exist uniquely at infinite coupling. 

By expanding the highest weight generating function of the Hilbert series at infinite coupling in powers of $q$, we have analysed the dressing of instanton operators by mesons and gauginos. For $N_f \leq5$ the defining equations for the space associated to the dressing can be obtained by keeping the relations at infinite coupling which are not corrected by the instanton operators. For $N_f=6,7$, the presence of charge $\pm$2 instanton operators as generators independent from the charge $\pm$1 ones complicates the picture and leaves the interpretation of the classical dressing in a preliminary and unsatisfactory stage. 

The techniques developed in this paper could also be applied to other 5d $\mathcal{N}=1$ theories with known Higgs branch at infinite coupling. We leave this to future work. The long term goal is to better understand supersymmetric instanton operators and their dressing from first principles and use such knowledge to derive a general formula for the Hilbert series associated to the Higgs branch at infinite coupling. We hope that the results of this paper can shine some light in this direction.

\acknowledgments

We are grateful to Nick Dorey, Kazuo Hosomichi, Ken Intriligator, Neil Lambert and Diego Rodriguez-Gomez for useful discussions.
S.C. and A.H. thank the Galileo Galilei Institute for Theoretical Physics (workshop ``Holographic Methods for Strongly Coupled Systems'') for hospitality and INFN for partial support during the completion of this work. 
A.H. would also like to thank Perimeter Institute and the Korea Institute for Advanced Studies for the kind hospitality received. 
 N.M. gratefully acknowledges the following institutes, workshops and researchers for their kind hospitality and partial support: Modern Developments in M-theory Workshop (South Korea), Kimyeong Lee and Seok Kim; Kavli IPMU, World Premier International Research Center Initiative (WPI Initiative), MEXT, Japan, Masahito Yamazaki and Tirasan Khandhawit; University of Torino and Carlo Angelantonj; University of Milano-Bicocca and Alberto Zaffaroni; ETH Zurich and Matthias Gaberdiel; LPTHE at the University Pierre et Marie Curie (Jussieu), Nick Halmagyi and Claudius Klare; Harvard University and Michele del Zotto; University of Rome Tor Vergata, Francesco Fucito, Francisco Morales and Massimo Bianchi. G.F. is supported by an STFC studentship.

\appendix 

\section{Hilbert series of chiral rings with gaugino superfields} \label{appSec:chiralRingHSwithGaugino}

In this appendix we present a method to compute the Hilbert series of the Higgs branch at finite coupling. In this computation we include the classical chiral operators as well as the gaugino superfield $\CW$.  

In five dimensions, the gaugino $\lambda^A_I$ carries the $USp(4)$ spin index $A=1,\ldots,4$ and the $SU(2)_R$ index $I= 1,2$. Since we focus on holomorphic functions, which are highest weights of $SU(2)_R$ representations, we restrict ourselves to $I=1$. In 4d $\CN=1$ language, which we adopt throughout the paper, the fundamental representation of $USp(4)$ decomposes to $[1;0]+[0;1]$ of $SU(2) \times SU(2)$.  These are usually denoted by undotted and dotted indices, respectively.  Since the latter correspond to non-chiral operators in the 4d $\CN=1$ holomorphic approach, we adhere to the undotted $SU(2)$ spinor index.  The gaugino superfield is henceforth denoted as ${\cal W}_\alpha$.

We will see that the 4d $\CN=1$ formalism adopted in this appendix yields results for the Hilbert series that are consistent with the chiral ring obtained by setting instanton and anti-instanton operators to zero in the five dimensional UV fixed point, which is discussed in the main body of the paper.

\subsection{$SU(2)$ gauge theory with $N_f$ flavours}

Let us denote the chiral matter fields appearing in the Lagrangian by $Q^i_a$, with $i=1, \ldots, 2N_f$ and $a=1,2$. 
The $F$-terms relevant to the classical Higgs branch are\footnote{Here and in the main body of the paper, our relations are valid in the chiral ring.  As operator relations, they hold up to a superderivative.}
\bea
\epsilon^{ab} \epsilon^{cd} Q^i_a  Q^i_d=0~. \label{rel1}
\eea
These relations are symmetric under the interchange of the indices $b$ and $c$.

Now let us discuss the inclusion of the gaugino superfield $({\cal W}_\alpha)_{ab}$.  ${\cal W}_\alpha$ is adjoint valued and is chosen to be a traceless symmetric 2-index tensor:
\bea
\epsilon^{ab} ({\cal W}_\alpha)_{ab} = 0 \label{rel2}~.
\eea
Moreover, we impose the following conditions (see section 2 of \cite{Cachazo:2002ry}) :
\bea
& \text{Each component of $({\cal W}_\alpha)_{ab}$ is an anti-commuting variable}~, \label{rel3} \\
& \epsilon^{bc} (\CW_\alpha)_{ab} (\CW_\beta)_{cd} + (\beta \leftrightarrow \alpha) =0 \quad \text{$\forall$ $\alpha,\beta=1,2$, $a,d=1,2$}, \label{rel4} \\
& \epsilon^{bc} (\CW_\alpha)_{ab} Q^i_c=0\quad \qquad \text{$\forall$ $\alpha=1,2$, $a=1,2$, $i=1,\ldots, N_f$}~. \label{rel5}
\eea
The condition \eref{rel3} follows from the fact that the lowest component of the gaugino superfield is fermionic.  The relation \eref{rel4} follows from gauge invariance and supersymmetry.  The relation \eref{rel5} indicates how the gaugino superfield acts on fundamental fields.
 
The Hilbert series of the ring of variables $Q^i_a$, $(\CW_\alpha)_{ab}$ subject to the conditions \eref{rel1}, \eref{rel2}, \eref{rel3}, \eref{rel4} and \eref{rel5} can be computed using {\tt Macaulay2}.
For reference, we provide the Macaulay2 code for the case of $N_f=3$ in source code (SC) \ref{sourcecode:codeM2Nf3}.

After integrating over the $SU(2)$ gauge group and restricting to the scalar sector under the Lorentz group, we obtain the Hilbert series of the space
\bea
\widetilde{\CM}_{1, SO(2N_f)} \cup \mathbb{Z}_2~,
\eea
where $\widetilde{\CM}_{1, SO(2N_f)}$ is the reduced moduli space of one $SO(2N_f)$ instanton on $\BC^2$ and $\mathbb{Z}_2$ is the moduli space generated by the glueball superfield $S$ such that $S^2=0$
\be
\begin{split}
H[ \widetilde{\CM}_{1, SO(2N_f)} \cup \mathbb{Z}_2 ](t; \vec x,w) &=  H[\mathbb{Z}_2](t;w)+ H[\widetilde{\CM}_{1, SO(2N_f)}] (t; \vec x) -1 \\
&= w^2 t^2 + \sum_{p=0}^\infty [0,p,0,\ldots,0] t^{2p}~,
\end{split}
\ee
where the fugacity $w$ counts the number of gaugino superfields $\CW$ and $\vec x$ are the fugacities of $SO(2N_f)$.  The plethystic logarithm up to order $t^4$ of this is
{\small
\be
\begin{split}
&\PL\left[H[ \widetilde{\CM}_{1, SO(2N_f)} \cup \mathbb{Z}_2 ](t; x,w) \right] = ([0,1,0,\ldots,0]+w^2)t^2 - \Big(1+[2,0,\ldots,0] \\
&\qquad\qquad\qquad\qquad\qquad\qquad +[0,0,0,1,0,\ldots,0] + w^2[0,1,0,\ldots,0] +w^4  \Big)t^4+\ldots~.
\end{split}
\ee}
This shows that the generators are the meson $M^{ij}=-M^{ji}$, in the adjoint representation of $SO(2N_f)$, 
and the glueball $S=-\frac{1}{32 \pi^2} \Tr \mathcal{W}_\alpha \mathcal{W}^\alpha$, subject to the relations
\bea
M^{ij}M^{jk}=0~, \quad M^{[ij}M^{kl]} =0~,\quad SM^{ij} = 0~, \quad S^2 = 0~.
\eea

\begin{sourcecode}[htbp]
\includegraphics[scale=0.5]{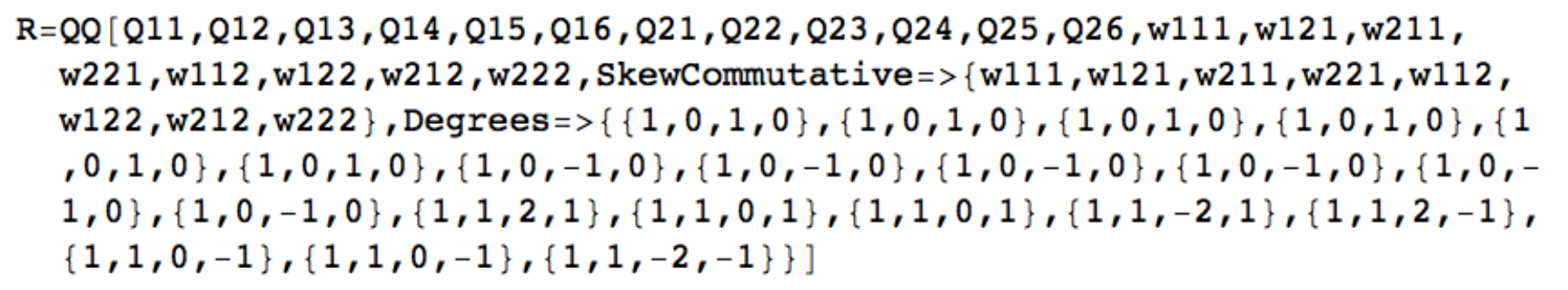}
\includegraphics[scale=0.5]{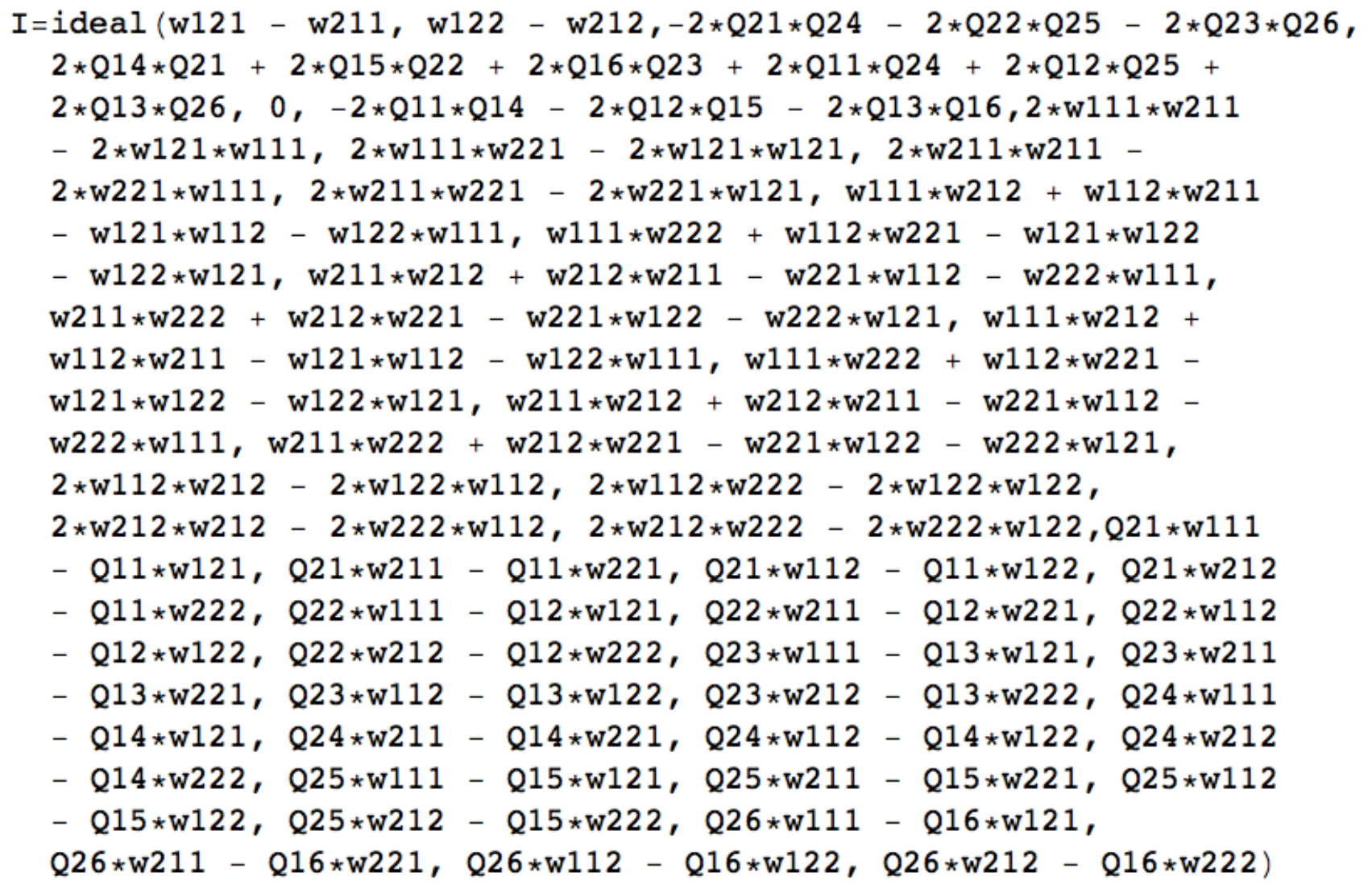}
\includegraphics[scale=0.5]{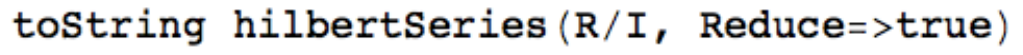}
\caption{A {\tt Macaulay2} code to compute the Hilbert series of the ring of variables $Q^i_a$, $(\CW_\alpha)_{ab}$, with $N_f=3$, subject to the conditions \eref{rel1}, \eref{rel2}, \eref{rel3}, \eref{rel4} and \eref{rel5}. Here we write $Q^i_a$ as {\tt Qai} and $(\CW_\alpha)_{ab}$ as {\tt wab}$\mathbf{\alpha}$.  The ring {\tt R} is multi-graded with respect to the following charges (in order): 1. the $R$-charge associated with the fugacity $t$, 2. the number of gaugino superfields associated with the fugacity $w$, 3. the weights of the $SU(2)$ gauge group, and 4. the weights of the $SU(2)$ symmetry associated with the index $\alpha$.}
\label{sourcecode:codeM2Nf3}
\end{sourcecode}

\subsection{$USp(2k)$ gauge theory with one antisymmetric hypermultiplet}

The analysis is similar to the previous subsection.  Let us denote the antisymmetric fields by $X^{ij}_a$, where $a=1,2$ and $i,j=1, \ldots, 2k$ are the $USp(2k)$ gauge indices. The $F$-terms associated to the classical Higgs branch is 
\bea
 J_{ii'}J_{jj'}J_{kk'} \epsilon^{ab} X^{ij}_{a}  X^{k' i'}_b =0~, \label{rel1b}
\eea
where $J_{ij}$ is the symplectic matrix associated with $USp(2k)$.

For the gaugino superfield $\CW^{ij}_\alpha$ (with $\alpha= 1,2$), we impose the conditions \cite{Cachazo:2002ry}
\bea
& {\cal W}^{ij}_\alpha = {\cal W}^{ji}_\alpha ~,\label{rel2b}\\
& \text{each component of ${\cal W}^{ij}_\alpha$ is an anti-commuting variable}~, \label{rel3b} \\
& J_{jk} \CW^{ij}_{(\alpha} \CW^{kl}_{\beta)}=0, \label{rel4b} \\
& J_{jk} (\CW^{ij}_\alpha  X^{kl}_a- X^{ij}_a  \CW^{kl}_\alpha)=0 ~. \label{rel5b}
\eea

After integrating over the $USp(2k)$ gauge group and restricting to the scalar sector under the Lorentz group, we obtain the Hilbert series of the space
\bea
\Sym^k \left( \BC^2 \times \mathbb{Z}_2 \right)~,
\eea
In particular, for $k=2$, we recover the Hilbert series \eref{USp41antisym}.

\section{$N_f=6$ in representations of $SO(12) \times SU(2)$}  \label{appSec:SO12xSU2}

Here we rewrite \eref{HSE7Nf6} and \eref{plE7SO12} in terms of characters of representations of $SO(12) \times SU(2)$:
\bea
& H[E_7](t; \vec x, y)  \nn \\
&= 1 + ([0, 0, 0, 0, 0, 0; 2] + [0, 0, 0, 0, 1, 0; 1] + 
    [0, 1, 0, 0, 0, 0; 0]) t^2  \nn \\
 & \quad + 
(1 + [0, 0, 0, 0, 0, 0; 4] + 
    [0, 0, 0, 0, 1, 0; 1] + [0, 0, 0, 0, 1, 0; 3] + 
    [0, 0, 0, 0, 2, 0; 2]  \nn \\
 & \quad + [0, 0, 0, 1, 0, 0; 0] + 
    [0, 1, 0, 0, 0, 0; 2] + [0, 1, 0, 0, 1, 0; 1] + 
    [0, 2, 0, 0, 0, 0; 0])  t^4  \nn \\
& \quad + \ldots~. \label{HSSO12SU2}
\eea
The plethystic logarithm of \eref{HSSO12SU2} is
\bea
&\PL \left[ H[E_7](t; \vec x, y) \right]  \nn \\
&=([0, 0, 0, 0, 0, 0; 2] + [0, 0, 0, 0, 1, 0; 1] + [0, 1, 0, 0, 0, 0; 0]) t^2-\Big(2  + [0, 0, 0, 1, 0, 0; 0]    \nn \\
& \quad + 
 [2, 0, 0, 0, 0, 0; 0]  +[0, 0, 0, 0, 1, 0; 1]+ [1, 0, 0, 0, 0, 1; 1] + 
 [0, 1, 0, 0, 0, 0; 2]  \Big) t^4 \nn \\
&\quad +\ldots~.
\eea

The representation $[0, 0, 0, 0, 0, 0; 2]$ corresponds to $I_{2+}$, $I_{2-}$ and $S$, $[0, 0, 0, 0, 1, 0; 1]$ to $I_{1 \pm}$ and $[0, 1, 0, 0, 0, 0; 0]$ to $M$.  In the Hilbert series \eref{HSSO12SU2} there is only one independent singlet at order $t^4$: this means that the singlets coming from these three sets of operators must be proportional to each other.  These indeed correspond to the trace part of \eref{rel1E7} and the relation \eref{rel2E7}.

\bibliographystyle{ytphys}
\bibliography{ref}

\end{document}